\documentclass[twocolumn,twocolappendix]{aastex631}

\usepackage{float}

\newcommand{\msigma}{\texorpdfstring{$\mathrm{M_{\bullet}}-\sigma\ $}{M-sigma}}
\newcommand{\mmstar}{\texorpdfstring{$\mathrm{M_{\bullet}}-\mathrm{M_*}\ $}{M-Mstar}}
\newcommand{\mstarsigma}{\texorpdfstring{$\mathrm{M_*}-\sigma\ $}{Mstar-sigma}}
\newcommand{\kms}{$\mathrm{km\ s^{-1}}$}

\begin{document}

\title{Signatures of BH seeding on the \msigma relation: Predictions from the \texttt{BRAHMA} simulations}

\author{Jonathan Kho}
\affil{University of Virginia 
530 McCormick Rd
Charlottesville, VA 22904, USA; \href{mailto:yja6qa@virginia.edu}{yja6qa@virginia.edu}}
\affil{Virginia Institute for Theoretical Astronomy, University of Virginia, Charlottesville, VA 22904, USA}
\affil{The NSF-Simons AI Institute for Cosmic Origins, USA}

\author{Aklant Kumar Bhowmick}
\affil{University of Virginia 
530 McCormick Rd
Charlottesville, VA 22904, USA; \href{mailto:yja6qa@virginia.edu}{yja6qa@virginia.edu}}
\affil{Virginia Institute for Theoretical Astronomy, University of Virginia, Charlottesville, VA 22904, USA}
\affil{The NSF-Simons AI Institute for Cosmic Origins, USA}

\author{Paul Torrey}
\affil{University of Virginia 
530 McCormick Rd
Charlottesville, VA 22904, USA; \href{mailto:yja6qa@virginia.edu}{yja6qa@virginia.edu}}
\affil{Virginia Institute for Theoretical Astronomy, University of Virginia, Charlottesville, VA 22904, USA}
\affil{The NSF-Simons AI Institute for Cosmic Origins, USA}

\author{Alex M. Garcia}
\affil{University of Virginia 
530 McCormick Rd
Charlottesville, VA 22904, USA; \href{mailto:yja6qa@virginia.edu}{yja6qa@virginia.edu}}
\affil{Virginia Institute for Theoretical Astronomy, University of Virginia, Charlottesville, VA 22904, USA}
\affil{The NSF-Simons AI Institute for Cosmic Origins, USA}

\author{Niusha Ahvazi}
\affil{University of Virginia 
530 McCormick Rd
Charlottesville, VA 22904, USA; \href{mailto:yja6qa@virginia.edu}{yja6qa@virginia.edu}}
\affil{Virginia Institute for Theoretical Astronomy, University of Virginia, Charlottesville, VA 22904, USA}
\affil{The NSF-Simons AI Institute for Cosmic Origins, USA}

\author{Laura Blecha}
\affil{Department of Physics, University of Florida, Gainesville, FL 32611, USA}

\author{Mark Vogelsberger}
\affil{Department of Physics, Kavli Institute for Astrophysics and Space Research, Massachusetts Institute of Technology, Cambridge, MA 02139, USA}

\begin{abstract}

The James Webb Space Telescope (JWST) has identified a large population of supermassive ($10^6$–$10^8~\mathrm{M}_\odot$) black holes (BHs) in the early universe ($z \sim 4$–$7$).
Current measurements suggest that many of these BHs exhibit higher BH-to-stellar mass ratios than local populations, opening a new window into the earliest stages of BH–galaxy coevolution and offering the potential to place tight constraints on BH seeding and growth in the early universe. 
In this work, we use the \texttt{BRAHMA} simulations to investigate the impact of BH seeding on the \msigma relation. 
These simulations adopt heavy $\sim10^5~\mathrm{M}_{\odot}$ seeds and systematically varied BH seeding models, resulting in distinct predictions for seed abundances. 
We find that different seed models lead to different normalizations of the \msigma relation at higher redshifts ($z > 2$) across all $\sigma$, and at low redshift for systems with low $\sigma$ ($50~\mathrm{km,s^{-1}} \lesssim \sigma \lesssim 80~\mathrm{km,s^{-1}}$).
The most lenient seed model also shows negligible evolution in the \msigma relation across redshift, while more restrictive models have substantially lower normalization on the \msigma relation for high $\sigma$~($\sim100$~\kms) at high redshifts, and evolve upward toward the local relation.
We demonstrate that the \msigma evolution is a direct consequence of merger-dominated BH growth in low mass galaxies~($\lesssim10^9~M_{\odot}$) and accretion dominated BH growth in high mass ($\gtrsim10^9~M_{\odot}$) galaxies. 
Furthermore, the scatter in  the \msigma relation is larger for the more restrictive models due to the inability of many BHs to grow significantly beyond their seed mass.

\end{abstract}


\section{Introduction} \label{sec:intro}

It is well established that nearly every massive galaxy contains a supermassive black hole (SMBH) ($\mathrm{\mathrm{M_{\bullet}}}>10^6\mathrm{M}_\odot$) in its nucleus \citep{1998AJ....115.2285M, 2008ARA&A..46..475H, 1995ARA&A..33..581K}. These SMBHs have masses that range from $\sim10^6$ to $10^{10} \mathrm{M}_\odot$, and are thus distinct from the stellar mass black holes (BH) that are created from the death of a massive star. While the path to forming a stellar mass BH is well-known \citep{1983bhwd.book.....S, 2023arXiv230409350H, 2025A&A...695A..71L}, it is still currently unknown how these SMBHs formed \citep{2010A&ARv..18..279V, 2020ARA&A..58...27I}.

Various different seeding channels have been proposed, which are distinguished by forming light, intermediate, and heavy seeds. Light seeds ($\mathrm{M_{seed}} \sim 10^2\mathrm{M}_\odot$) are expected to be the remnants of some Population III stars, which are thought to have been ubiquitous in the early universe \citep{2023ARA&A..61...65K}. 
Their low masses, however, have been shown to make growth into the observed $10^{9-10}\rm{M}_\odot$ BHs very challenging \citep{2009ApJ...696L.146M, 2005ApJ...633..624V, 2016MNRAS.458.3047P, 2023ARA&A..61...65K, 2018MNRAS.480.3762S, 2009ApJ...696.1798T,2023MNRAS.524..176J,2016MNRAS.457.3356V}. 
Intermediate mass seeds ($\mathrm{M_{seed}}\sim10^{3-4}\mathrm{M}_\odot$) may be formed via stellar collisions or hierarchical BH mergers in a dense nuclear star cluster \citep{2002ApJ...576..899P,2023MNRAS.526..429A,2023MNRAS.521.3553R, 2022ApJ...927..231F}. Heavy seeds ($\mathrm{M_{seed}}\sim10^{4-5}\mathrm{M}_\odot$), thought to be formed through the direct collapse of a cloud of gas, are much more difficult to form but have the benefit of being able to grow more easily due to their larger initial seed mass. 
In particular, producing a heavy seed requires environmental conditions like low metallicities or a critical Lyman-Werner flux which suggests that they are too rare to explain the nunmber densities of either the high-$z$ AGN or the local SMBH populations \citep{2024OJAp....7E..72R}.

Beyond having unknown origins, these SMBHs also curiously have very tight scaling relations with properties of their host galaxies ~\citep[see][for a review]{2013ARA&A..51..511K}. Specifically, in the local Universe, we find that the measured masses~($\mathrm{M_{\bullet}}$) of SMBHs strongly correlate with the stellar mass~($\mathrm{M_*}$) and stellar velocity dispersion of the host galaxy bulge~($\sigma$) \citep{2015ApJ...813...82R, 1998AJ....115.2285M, 2003ApJ...589L..21M, 2000ApJ...539L..13G, 2000ApJ...539L...9F}. Both of these correlations are remarkable because the spatial extent of a galaxy ($\sim$kpc) far exceeds the region directly influenced by the SMBH’s gravity ($\sim$pc). These correlations could suggest a ‘coupled’ evolution between BHs and their host galaxies, or they might arise as a statistical consequence of the concurrent growth of black holes and galaxies through hierarchical structure formation. The coupled evolution of BHs and host galaxies as a consequence of AGN feedback has been studied at length in simulations, and suggests that AGN activity may play an important role in driving these tight correlations \citep[e.g.][]{2019MNRAS.484.4413H,2020MNRAS.493.1888T,2005Natur.433..604D}.

In the new era of the James Webb Space Telescope, our understanding of both BH seeding models and BH-galaxy scaling relations is being transformed as a population of high-$z$ ($z=4-7$) broad line AGN (BLAGN) with masses of $\mathrm{M_{\bullet}} \sim 10^{6-8}\mathrm{M}_\odot$ has been discovered \citep{2024Natur.627...59M, 2024ApJ...960L...1N, 2024NatAs...8..126B, 2024ApJ...964...39G, 2024ApJ...974..147L, 2023ApJ...959...39H, 2024ApJ...965L..21K}. 
The discovery of these AGN allows us to probe galaxy scaling relations at high redshift for the first time.
This emerging population of BHs appears to be $10-100$ times more massive than what would be predicted by local stellar mass scaling relations, suggesting that the \mmstar relation may evolve with redshift \citep{2023ApJ...957L...3P,  2024ApJ...964..154P, 2024A&A...691A.145M, 2025arXiv250303675S, 2024Natur.628...57F}. 
The ubiquity of these JWST BLAGN \citep{2023ApJ...959...39H, 2024arXiv240610341A, 2024ApJ...968...38K, 2024ApJ...964...39G}  that appear overmassive suggests that heavy BH seeds may not be as rare as previously thought, as overmassive BHs are one of the key signatures of heavy seeds \citep{2013MNRAS.432.3438A, 2017ApJ...838..117N, 2018ApJ...865L...9V, 2023MNRAS.519.2155S, 2024ApJ...960L...1N, 10.1093/mnras/stae1449}. 
While there does exist some debate as to whether a subpopulation of these BLAGN (dubbed little red dots (LRDs) due to their compact morphology and steep red rest-frame continuum) are truly reddened or dust-obscured AGN as opposed to dusty starburst galaxies \citep{2023ApJ...956...61A, 2024ApJ...969L..18A, 2024ApJ...968....4P, 2024ApJ...968...34W}, more recent work has seemed to strongly favor the AGN hypothesis, though several different AGN interpretations abound \citep{2025arXiv250316595R, 2025arXiv250316596N, 2025arXiv250403551J, 2025arXiv250506965K}. 
Nonetheless, these BH masses taken at face value pose strong constraints for BH seeding and growth \citep[see section 11 of][for a brief review]{2025arXiv250117078S}. 

Interestingly, \cite{2024MNRAS.531.4311B, 2024MNRAS.533.1907B} found that BHs seeded both as extrapolated seed descendants of light ($\mathrm{M_{seed}} = 10^3\mathrm{M}_\odot$) seeds and directly as heavy ($\mathrm{M_{seed}} = 10^5 \mathrm{M}_\odot$) seeds were over massive compared to the local \mmstar relation at $z=4-7$, in good agreement with the high-$z$ JWST AGN that have been discovered. 
\cite{2024MNRAS.533.1907B} also found that the \mmstar relation can provide strong constraints on BH seeding, since an overmassive \mmstar relation at high-$z$ suggests a more efficient production of heavy seeds than is typically assumed possible. 
Furthermore, while both \cite{2024A&A...691A.145M} and \cite{2025arXiv250403551J} found their sample of AGN to be over massive on the \mmstar plane, these same BHs were largely consistent with the local \msigma relation, leading to questions about which scaling relations evolve over cosmic time and why. 
Since \cite{2024MNRAS.533.1907B} found that signatures of BH seeding were found in the \mmstar relation due to different regimes of merger-driven versus accretion-driven BH growth, it is natural to search for these signatures in the \msigma relation as well.

\cite{2009MNRAS.400.1911V} used a Semi-Analytic model (SAM) to investigate potential signatures of black hole seeding in the \msigma relation. 
They found that massive seeds tend to form over massive relative to the local \msigma relation, and track horizontally onto it by growing in $\sigma$, whereas light seeds form under massive relative to the local relation and track upward  by growing the central BH mass while leaving $\sigma$ relatively unchanged. 
However, in their model for BH growth, they assumed the local \msigma relation to be constant across redshift, and were thus unable to form conclusions about a \msigma redshift evolution. 
While SAMs are useful, cheap models that can provide insights into the questions of BH seeding and its impact on galaxy scaling relations, cosmological simulations are necessary as they provide a more reliable and self-consistent method of tracking the seeding conditions, growth, and (co)-evolution of BHs and their host galaxies.

In light of previous work, our goal here is to search for signatures of BH seeding in the \texttt{BRAHMA} simulations by analyzing the slope, scatter, and possible redshift evolution of the \msigma relation for different BH seed models. 
Although other studies have examined the impact of different BH seeding prescriptions on (heavy) direct-collapse BH formation sites \citep[e.g.][]{2018ApJ...861...39D, 2020MNRAS.492.4917L}, and there have been many SAM studies examining the effects of different BH seeding prescriptions \citep[e.g.][]{2009MNRAS.400.1911V, 2023MNRAS.518.4672S, 2025MNRAS.536.2783E}, we are unaware of any other study thus far examining the effect of varying BH seeding prescriptions on the \msigma relation with cosmological simulations. 
The \texttt{BRAHMA} simulations are thus ideal for this work because they provide unique seeding criteria for heavy BH seeds that have not been implemented in other hydrodynamic simulations, and which have been studied in depth with cosmological zoom simulations prior to their implementation in these full volume simulations \citep{2021MNRAS.507.2012B,2022MNRAS.510..177B,2022MNRAS.516..138B}. 

The structure of this paper is as follows: we  describe the \texttt{BRAHMA} simulations and our methods in section \ref{sec: Methods}, present our results in section \ref{sec: Results}, comment on JWST observations and uncertainties in our results in section \ref{sec: Discussion} and provide concluding thoughts in section \ref{sec: Conclusions}.

\section{Methods} \label{sec: Methods}

The \texttt{BRAHMA} simulations were run using the AREPO gravity $+$ magnetohydrodynamic (MHD) solver \citep{2010MNRAS.401..791S, 2011MNRAS.418.1392P, 2016MNRAS.462.2603P, 2020ApJS..248...32W}. 
The gravity solver uses the Particle Mesh Tree \citep{1986Natur.324..446B} and the evolution of the gas is described by the ideal MHD equations solved over a dynamic unstructured grid generated via a Voronoi tessellation of the domain. 
All the simulations are characterized by the Planck Collaboration XIII (\citeyear{2016A&A...594A..13P}) cosmology i.e. $\Omega_\Lambda$ = 0.6911, $\Omega_\mathrm{m}$ = 0.3089, $\Omega_\mathrm{b}$ = 0.0486, $H_0 = 67.76\ \mathrm{km\ s^{-1}\ Mpc^{-1}}$, $\sigma_8$ = 0.8159, $n_s$ = 0.9667. 
Halos are identified using the Friends-of-Friends (FOF) algorithm \citep{1985ApJ...292..371D} with a linking length of 0.2 times the mean particle separation. 
Subhalos are identified using the \texttt{SUBFIND} \citep{2001MNRAS.328..726S} algorithm. The initial conditions were created using the \texttt{MUSIC} algorithm at $z=127$ \citep{2011MNRAS.415.2101H}. 
The simulations are run in a comoving volume of $[18~\mathrm{Mpc}]^3$ with $512$ dark matter (DM) particles. 
The DM mass resolution is $1.5\times 10^6 \mathrm{M}_\odot$, and the target baryon mass resolution is $2.4\times10^5\mathrm{M}_\odot$.

\subsection{Galaxy formation model}

Aside from BH seeding, the \texttt{BRAHMA} simulations adopt all of the features of the \texttt{IllustrisTNG} model
\citep[hereafter \texttt{TNG};][]{2018MNRAS.473.4077P, 2018MNRAS.479.4056W}, which itself is based on the \texttt{Illustris} model~\citep{2013MNRAS.436.3031V, 2014MNRAS.438.1985T}.
Radiative cooling includes the contributions from primordial species ($\mathrm{H, H^+, He, He^+, He^{++}}$ according to \citealt{1996ApJS..105...19K}) and metals~\citep{2009MNRAS.393...99W}. 
Metal cooling rates are interpolated from tables calculated 
in the presence of a spatially uniform, time-dependent UV background. 
As gas cools and condenses, star formation occurs when gas densities exceed $0.13\ \mathrm{cm}^{-3}$. 
Star-forming gas cells (i.e. those exceeding the star formation density threshold) are considered to represent a multi-phase ISM via a subgrid, effective equation of state \citep{2003MNRAS.339..289S}. 
Star particles represent full stellar populations with each particle being characterized by a single stellar age and metallicity. 
The initial stellar mass function is taken from \cite{2003PASP..115..763C}, and the stellar evolution model is based on \cite{2013MNRAS.436.3031V}  with modifications for \texttt{TNG} described in \cite{2018MNRAS.473.4077P}. 
Stellar chemical enrichment is modeled by following the evolution of (H, He, C, N, O, Ne, Mg, Si, Fe). 
Supernova feedback is modeled as a galactic-scale wind depositing mass, momentum, and metals to surrounding gas cells.

BHs in our simulations grow through two channels: accretion and mergers. 
Accretion is modeled with an Eddington-limited Bondi-Hoyle model \citep{1944MNRAS.104..273B, 1952MNRAS.112..195B}. 
The Eddington limit and BH bolometric luminosities are calculated with an assumed radiative efficiency of $\epsilon=0.2$. 
AGN feedback is modeled as one of two modes: `kinetic' or `thermal'. 
Thermal feedback is implemented for high Eddington ratios, wherein a fraction of the radiated luminosity is deposited in neighboring gas cells. 
Kinetic feedback is used for low Eddington ratios, wherein kinetic energy is injected to the gas in random directions at irregular time intervals. 
In our simulations, a BH merger occurs when at least one BH lies within the `neighbor search radius' of the other.  
Furthermore, BHs are re-positioned to the nearest potential minimum within 1000 nearest gas cells at each time step to avoid spurious gravitational kicks from massive DM particles, which are $\sim 10$ times more massive than our BH seeds. 
This means every halo merger wherein both halos have been seeded will also result in a prompt BH merger, which represents a highly optimistic merger scenario. 
The effects of a more realistic inspiral via the effects of dynamical friction are investigated in a future work by Bhowmick et al. (in prep). 
 
For further details regarding the models used here, please see \cite{2024MNRAS.533.1907B} and references therein.

\subsection{Black hole seeding prescriptions}\label{sec: BH_seeding}

The key feature distinguishing \texttt{BRAHMA} from \texttt{TNG} is the prescriptions for seeding BHs within subhalos. 
In this study, we use four simulations that include increasingly restrictive seed models by stacking the seeding requirements described below.
In all four boxes, BHs are seeded with an initial seed mass of $\mathrm{M_{seed}}=1.5 \times 10^5 \mathrm{M}_\odot$. 
The criteria imposed are motivated based on conditions expected for direct collapse black hole (DCBH) formation and are as follows:

\begin{enumerate}
    \item \textit{Halo mass criterion}: BHs can only be seeded within subhalos that are resolved. The requirement for \texttt{BRAHMA} subhalos to be resolved is 32 DM particles, which corresponds to a halo mass of $4.8\times10^7h^{-1}\mathrm{M}_\odot$.

    \item \textit{Dense and metal-poor gas mass criterion}: Seeds are placed in halos that exceed a minimum amount of gas mass~($5\times \mathrm{M_{seed}}$) in local pockets of dense ($\rho>0.13\ \mathrm{cm}^{-3}$) and metal-poor ($Z < 10^{-4}Z_\odot$) gas.
    
    \item \textit{Lyman-Werner flux criterion}: \
    The dense and metal-poor gas is additionally required to be exposed to an incident Lyman-Werner~(LW) flux greater than $\rm{J}_{\rm crit} =  10\ \mathrm{J}_{21}$. 
    The critical LW flux~($\rm{J}_{\rm crit}$) used here is significantly lower than values suggested by small scale simulations and one-zone chemistry models~\citep{2010MNRAS.402.1249S,2014MNRAS.445..544S}, who report a necessary intensity of $\geq1000\ \mathrm{J}_{21}$. But as shown in \cite{2024MNRAS.533.1907B}, to produce overmassive BHs consistent with JWST, we need more abundant heavy seed formation. Therefore, we used a low value of  $\mathrm{J}_{21}$ that has been shown to be feasible if gas is dynamically heated during major mergers~\citep{2020OJAp....3E..15R, 2020MNRAS.492.3021R, 2019Natur.566...85W}. 
    
    \item \textit{Gas spin criterion}: With the addition of this criteria, halos must have a sufficiently low spin parameter, defined by
    \begin{equation}
        \lambda = \frac{|\vec{J}_{\mathrm{spin}}|}{\sqrt{2}M_{\mathrm{gas}}R_{\mathrm{vir}}V_{\mathrm{vir}} }<\lambda_{\mathrm{max}},
    \end{equation}
    where $\vec{J}_{\mathrm{spin}}$ is the spin of the gas in dimensionless units, $M_{\mathrm{gas}}$, $R_{\mathrm{vir}}$, and $V_{\mathrm{vir}}$ are the gas mass, virial radius, and circular velocity, and $\lambda_{\mathrm{max}}$ is at the onset of the Toomre instability respectively \citep{1964ApJ...139.1217T}. This stability criterion was based on the analysis done by \cite{2006MNRAS.371.1813L}, who studied the stability of pre-galactic disks and the conditions necessary for the disk's collapse into a DCBH. 
    
    \item \textit{Halo environment criterion}: This criterion ensures that seeds are placed in halos that have a ``rich" environment. This means that the halo must have at least one halo of comparable or greater mass within five times its virial radius. This criterion is designed to seed exclusively in environments where halos are more likely to undergo major mergers that could dynamically heat gas, thereby justifying the feasibility of our adopted low $J_{\rm crit}$ value. 
\end{enumerate}

For more details on the implementation of the Lyman-Werner flux/gas spin and halo environment criteria, please see \cite{2022MNRAS.510..177B} and \cite{2024MNRAS.529.3768B}, respectively.
While all four simulations employ the halo mass criterion, the first simulation, which we refer to as BI, only additionally employs the dense and metal-poor criterion. 
The second simulation, BII, includes both the dense and metal-poor criteria as well as the Lyman-Werner flux criteria. 
The third simulation, BIII, includes both of the previous criteria, as well as the low spin criterion. 
Finally, the fourth simulation, which adds the rich environment criterion, we refer to as BIV. Note that in \cite{2024MNRAS.533.1907B} these same simulations were referred to as SM5, SM5\_LW10, SM5\_LW10\_spin, and SM5\_LW10\_spin\_rich respectively.

In order to understand the effects that these above ``gas based" seeding criterion have on the scaling relations, we compare our simulation results to the scaling relations for the \texttt{TNG} \citep{2018MNRAS.475..624N, 2018MNRAS.475..648P, 2018MNRAS.477.1206N, 2018MNRAS.480.5113M, 2018MNRAS.475..676S} simulation that uses the standard ``halo mass threshold-based" seeding. 
BHs are seeded with mass  $8\times10^5h^{-1} \mathrm{M}_\odot$ in \texttt{TNG}  in halos with masses greater than $5\times10^{10}h^{-1}\mathrm{M}_\odot$ \citep{2017MNRAS.465.3291W, 2013MNRAS.436.3031V}. 
This condition is significantly more restrictive than \texttt{BRAHMA}'s minimum halo mass requirement of $4.8\times10^7h^{-1}\mathrm{M}_\odot$. 

\begin{figure}
    \centering
    \includegraphics[width=1.0\linewidth]{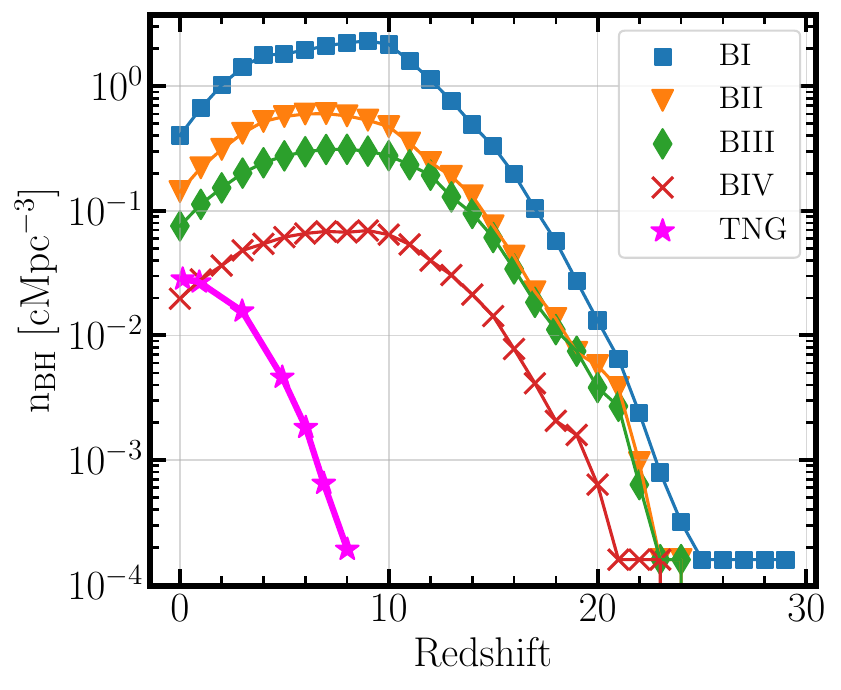}
    \caption{BH number density in comoving $\mathrm{Mpc^{-3}}$ for each of the four \texttt{BRAHMA} simulations as a function of redshift. With each incrementally more restrictive seed model, the number density of BHs decreases at all redshifts, and halos start to be seeded at later times. The onset of seeding and number density of BHs in \texttt{TNG} is greatly reduced due to its extremely restrictive halo mass requirement. Since the more restrictive seed models have much fewer BH seeds, their growth via mergers is significantly reduced.}
    \label{fig:BH_n}
\end{figure}

To demonstrate the impact of the various seeding prescriptions, we show the redshift evolution of the comoving BH number density for our four seeding models in Fig. \ref{fig:BH_n}. 
As one might expect, imposing increasingly restrictive seeding criteria reduces the number density of BHs in each subsequent simulation at all redshifts. 
Increasingly restrictive seeding criteria also delay the onset of seeding to somewhat later redshifts.
All four curves show a characteristic maximum number density at redshifts of approximately $z\sim6-10$, which is due to a balance between BH seeding and BH mergers, as well as metal enrichment at low-$z$. 
At early times, the BH seeding rate is much greater than the merger rate of seeded halos (since very few halos have yet to be seeded), causing the number density to increase. 
As halos continue to be seeded in increasing numbers until $z\lesssim10$ and then merge, the reduction in the BH number densities due to mergers greatly increases, resulting in a maximum number density followed by a gradual decline for all four simulations.
For the \texttt{TNG} simulation, plotted for comparison, the onset of seeding and peak number density is greatly delayed due to the extremely restrictive halo mass criterion. 
Because of the simple seeding criterion, the number of halos being seeded in \texttt{TNG} still increases at later times ($z<10$).

\subsection{The \texorpdfstring{$\sigma$} \ calculation}

We calculate the velocity dispersion as the mass-weighted velocity dispersion of all of the stars in the bulge of the galaxy. 
To identify the stars that belong to the bulge, we perform a kinematic decomposition as done in \cite{2018MNRAS.478.5063H} and \cite{2003ApJ...591..499A}. 
A metric $\kappa = j/j_{\mathrm{max}}(e)$ is calculated for every star in the galaxy, where $j$ is the star's specific angular momentum, and $j_{\mathrm{max}}(e)$ is the maximum specific angular momentum for a star with specific binding energy $e$. 
The specific binding energy of a star particle is defined as the energy required to make the star an unbound particle, and was calculated as the sum of the specific kinetic and potential energies. 
Stars with $\kappa>0.5$ are considered to be a part of the disk of the galaxy, in following with \cite{2019ApJ...884..129D,2020ApJ...895..139D}. 
Stars with $\kappa<0.5$ are considered a part of the bulge, and used in our stellar mass and $\sigma$ calculations. 
To allow for a proper comparison between our $\sigma$ calculations and observed line-of-sight $\sigma$'s, in all of our plots we show a one dimensional $\sigma$ calculated as $\sigma_{\mathrm{1D}} = \frac{\sigma_{\mathrm{3D}}}{\sqrt{3}}.$ 
Here, $\sigma_{\mathrm{3D}} = \sqrt{\sigma_x^2 + \sigma_y^2 +\sigma_z^2}$, where $\sigma_x,\sigma_y,\sigma_z$ are the stellar velocity dispersions in the $x$, $y$, and $z$ directions respectively. 

In keeping with \cite{2018MNRAS.478.5063H}, we only retain subhalos with greater than 1000 stars in our sample for \texttt{TNG}, as the decomposition may not produce well-defined bulge and disk components below this number. 
However, we apply the decomposition indiscriminately to all subhalos in the \texttt{BRAHMA} simulations, as imposing a 1000 star criterion would eliminate most, if not all, subhalos in our simulations at high redshift. 
We thus apply the decomposition to all \texttt{BRAHMA} subhalos for each redshift, regardless of the number of stars in each subhalo.
The decomposition is performed based on the energetics of individual star particles relative to their gravitational potential environment.
Each star particle is classified as being on a `bulge-like' or `disk-like' orbit to recover the spheroidal component of lower-mass galaxies.
We note that while this method is imperfect, it yields reasonable cuts for the bulge component of most galaxies in our sample.
It is also worth noting that while we perform this decomposition to have bulge stellar masses and  velocity dispersions to allow for comparison to the observed scaling relations, photometrically derived bulge-disk decompositions may differ from kinematically derived decompositions, especially for lower stellar mass systems \citep[see Fig. 7 of][]{2017MNRAS.467.2879B}.

The scripts used to perform this decomposition and the rest of the analysis in this work are publicly available at \url{https://github.com/khojonat/Brahma_Analysis}.

\section{Results} \label{sec: Results}

\subsection{Impact of seeding on the \msigma relation}\label{sec: seeding_impact}

\begin{figure*}
    \centering
    \includegraphics[width=1.0\linewidth]{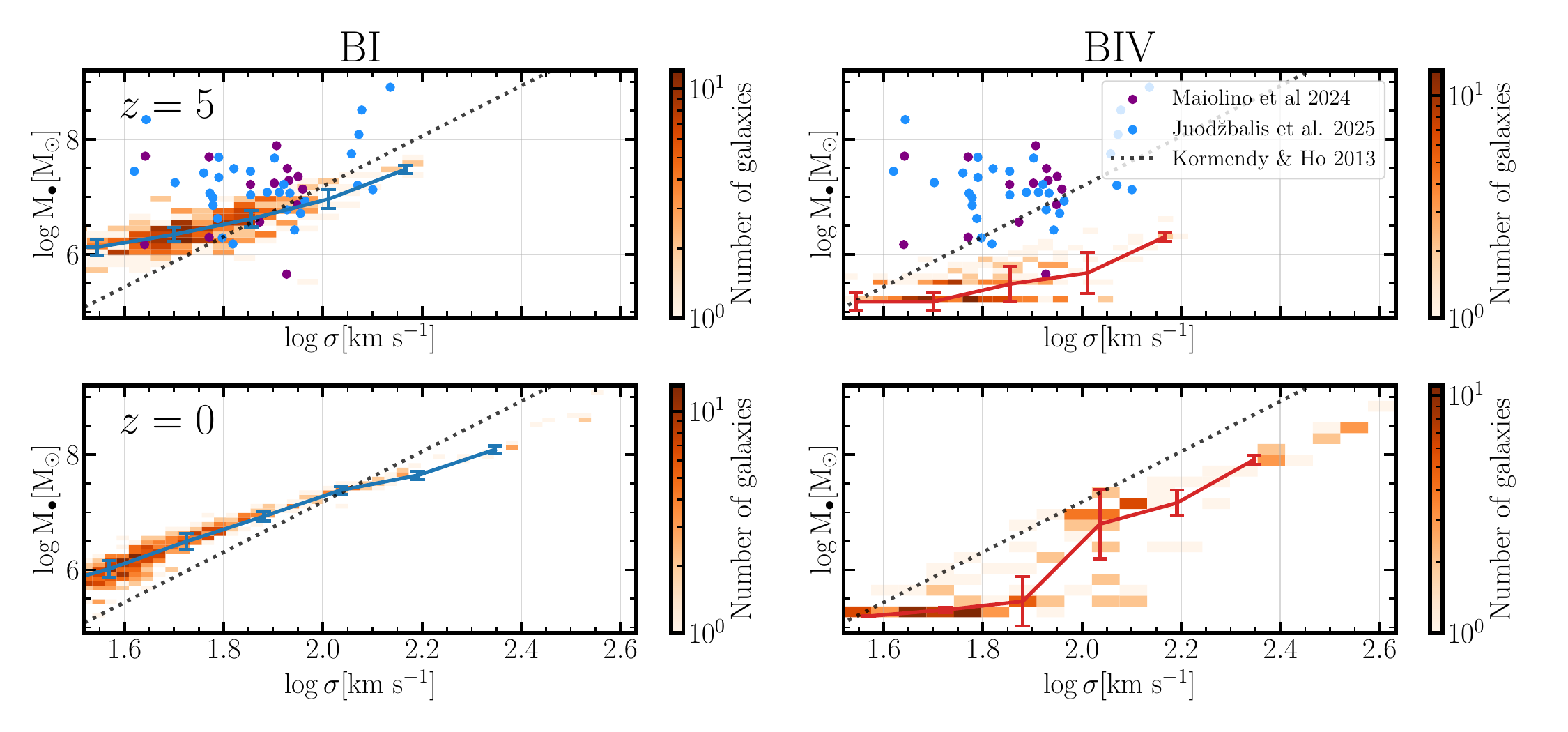}
    \caption{Histograms of the full scatter of BHs for the most lenient and most restrictive seed models on the \msigma relation. Plotted in blue and red is the median trend of the data, with error bars representing the interquartile range in BH mass. The dotted line shows the observed \msigma by KH13. While the relation retains a tight correlation for the most lenient model across redshift, the scatter drastically increases with decreasing redshift for the most strict model at intermediate values of $\sigma$.}
\label{fig:Fullscatter}

    \centering
    \includegraphics[width=0.9\linewidth]{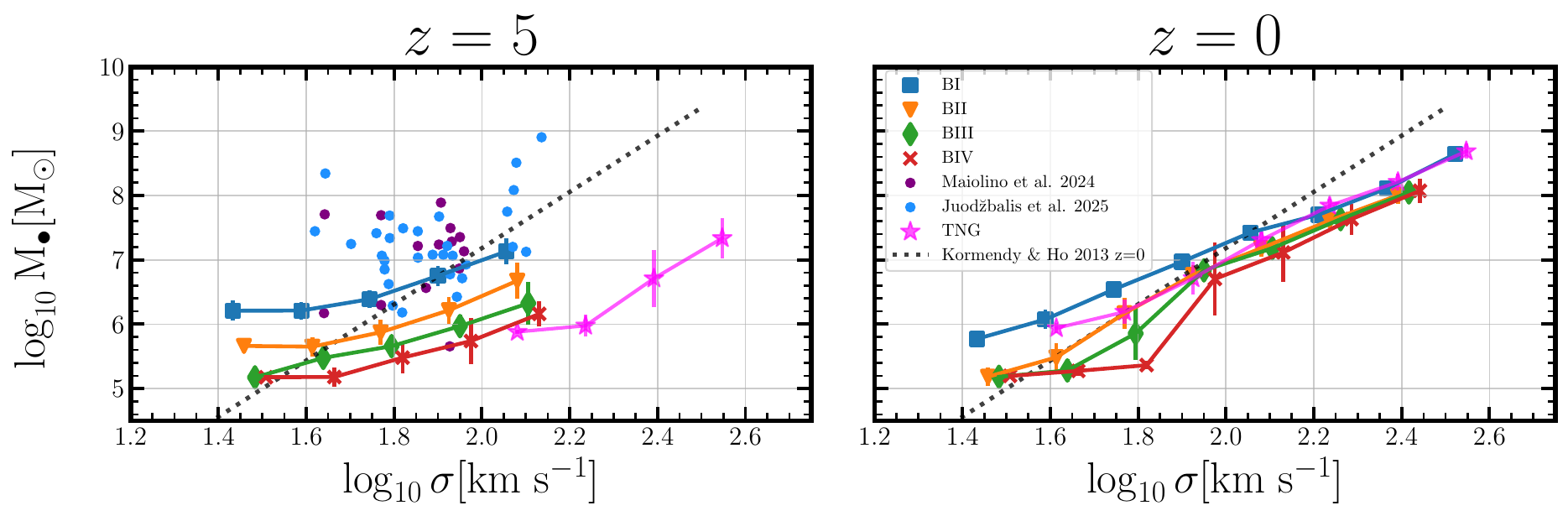}
    \caption{Comparison of the \texttt{BRAHMA} and \texttt{TNG} simulations' \msigma relations at $z=5, 0$. The local relation found by KH13 is plotted in both panels for reference, and high-$z$ AGN candidates found by \protect\cite{2024A&A...691A.145M} and \protect\cite{2025arXiv250403551J} are shown in purple and blue in the $z=5$ panel. With the exception of our most lenient seed model, the simulations all begin under-massive to the local relation at high-$z$ and converge to this relation by $z=0$. The different \texttt{BRAHMA} seed models are clearly differentiable for all $\sigma$'s at $z=5$ via their different normalizations, and are still differentiable for $10^{1.7}$ \kms$ \lesssim\sigma\lesssim$ $10^{1.9}$ \kms at $z=0$.}
    \label{fig:msigmabinned}

\end{figure*}

We begin by showing in Fig. \ref{fig:Fullscatter} 2D histograms of the \msigma relation for the most lenient~(BI) and most strict~(BIV) seeding models at redshifts $z=5, 0$. 
The blue and red lines trace the median BH mass and interquartile range for different $\sigma$ bins. In all of our figures, we also include the local scaling relations for \msigma and \mmstar as found by \citealt{2013ARA&A..51..511K} (hereafter KH13), the high-$z$ \mmstar relation found by \cite{2023ApJ...957L...3P} (hereafter, P23), and a population of recently detected high-redshift AGN as reported by \cite{2024A&A...691A.145M} and \cite{2025arXiv250403551J}. 
We will first discuss the trends exhibited by the different simulations in this section, followed by how they the compare against observations in Section \ref{sec: Observations}.

Fig. \ref{fig:Fullscatter} shows the range of velocity dispersions that our boxes can probe at different redshifts, which is naturally limited by our volume and resolution. 
At $z=5$, our \texttt{BRAHMA} boxes are able to probe $\sigma$ values between $\sim 10^{1.5} - 10^{2.2}~\rm km~s^{-1}$, whereas at $z=0$ they probe $\sim10^{1.5}-10^{2.6}~\rm km~s^{-1}$.

Most interestingly, we can clearly see that the BH masses are substantially different between the seed models, thereby impacting the median as well as the scatter of the \msigma relation. 
At $z=5$, we find that the most lenient BI model produces a substantially higher \msigma normalization~(higher BH mass at fixed $\sigma$) compared to the most restrictive BIV model. 
In the left panel of Figure \ref{fig:msigmabinned}, we can see this trend more readily as the median trends for all four seed models are plotted together. 
This is because in the more lenient seed models, there is stronger merger-driven BH growth due to the higher number of seeds. 
To that end, \cite{2024MNRAS.533.1907B} and \cite{2025MNRAS.538..518B} showed that at these high redshifts, the BH mass assembly is dominated by mergers in these simulations. 
In fact, \texttt{TNG}, which has the most restrictive seed model, produces a lower \msigma than all the \texttt{BRAHMA} boxes~(albeit \texttt{TNG} probes much higher $\sigma$ values due to lower resolution and a larger volume). 
This is simply because too few seeds are produced in \texttt{TNG} to fuel enough merger-driven growth.

By $z=0$, the \msigma relations show substantially reduced seed model variations, particularly for $\sigma \gtrsim 80~\rm km~s^{-1}$ (bottom panels of Figure \ref{fig:Fullscatter} and right panel of Figure \ref{fig:msigmabinned}). 
These high $\sigma$ values correspond to more massive galaxies, where \cite{2025MNRAS.538..518B} found that BH mass assembly is dominated by gas accretion, in line with the expectations from the Soltan Argument ~\citep{1982MNRAS.200..115S}. 
However, seed model variations are much more substantial at low $\sigma$ values between $50$ to $80~\rm km~s^{-1}$. 
This is because for these low $\sigma$ values corresponding to lower mass~($\lesssim 10^9~M_{\odot}$) galaxies, \cite{2025MNRAS.538..518B} showed that BH growth continues to be dominated by mergers all the way down to $z=0$.

Overall, we find that the median \msigma relations show the strongest seed model variations at $z=5$ across all $\sigma$ values probed by our simulations, and at $z=0$ for low $\sigma$ values~($\sim 50$–$80$ \kms). 
These variations primarily arise from differences in the ability to fuel merger-driven BH growth under the various seeding models.

\subsection{Redshift evolution of the median \msigma relation: Relationship to \mmstar}
\label{sec: MEGA}

\begin{figure*}
    \centering
    \includegraphics[width=0.8\linewidth]{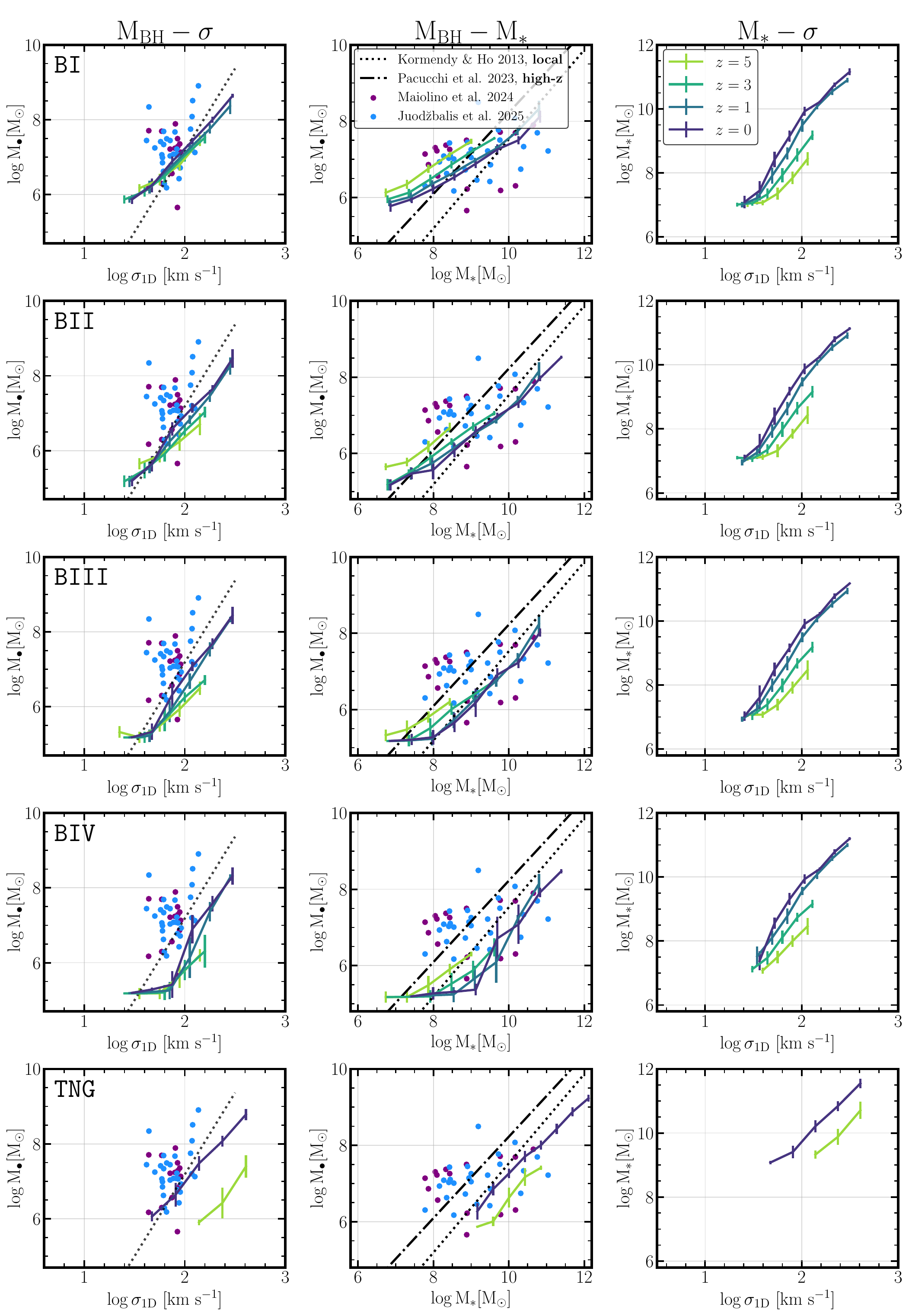}
    \caption{The \msigma, \mmstar, and \mstarsigma relations for the four \texttt{BRAHMA} boxes and \texttt{TNG}. Each plot shows the relation for redshifts $z = 5,3,1,0$ to provide a sense of their redshift evolution. The \msigma and \mmstar columns show the local relations as reported by KH13, and the \mmstar plots additionally show the high-$z$ inferred relation found by P23. These columns also show high-$z$ AGN as reported by \protect\cite{2024A&A...691A.145M} and \protect\cite{2025arXiv250403551J}.}
    \label{fig:MEGA}
\end{figure*}

In this section, we delve deeper into the redshift evolution of the median \msigma relation for the different seed models and how it connects to the \mmstar and \mstarsigma relations.
Fig. \ref{fig:MEGA} simultaneously shows the three scaling relations (\msigma, \mmstar, and \mstarsigma) for each of our \texttt{BRAHMA} boxes, as well as for \texttt{TNG}.
Notably, for the most lenient seed model, the median \msigma relation~(1st row, left panel) is largely insensitive to redshift. 
For the more restrictive seed models, the \msigma evolution follows a ``bottom-up" trend at high $\sigma$~($\gtrsim100$ \kms) i.e. BH masses are lower at earlier times. 
The evolution is much weaker at the low $\sigma$ end. 
Continuing this trend, the \texttt{TNG} seed model, being even more restrictive than BIV, produces a bottom-up \msigma evolution which is even more pronounced than the \texttt{BRAHMA} boxes.

While the \msigma evolution proceeds in a bottom-up fashion for most of our models, the \mmstar relation (middle column of Fig. \ref{fig:MEGA}) exhibits a 'top-down' evolution across all \texttt{BRAHMA} boxes—that is, BH masses are higher at earlier times at fixed stellar mass. 
This evolution is seen primarily for galaxies in the $\sim10^7$–$10^9 \rm{M}_{\odot}$ range, as more massive systems are not accessible in our \texttt{BRAHMA} volumes at higher redshifts. 
As shown in \cite{2025MNRAS.538..518B}, BH growth in these relatively low-mass galaxies is dominated by mergers down to $z \sim 0$. 
In such systems, the top-down evolution of the \mmstar relation arises naturally from this merger-driven BH growth: while BHs grow mainly via mergers, stellar mass increases more rapidly due to a combination of in-situ star formation and mergers. 
Notably, in contrast to the top-down \mmstar evolution seen in \texttt{BRAHMA}, the \texttt{TNG} relation (Fig. \ref{fig:MEGA}, bottom row) shows a bottom-up evolution. 
This is due to (1) a significantly lower BH seed abundance in low-mass galaxies, which suppresses merger-driven BH growth at $\rm{M}_* \lesssim 10^9 \rm{M}_{\odot}$, and (2) the fact that we show only galaxies with $\rm{M}_* \gtrsim 10^9~\rm{M}_{\odot}$ (due to the 1000 star-particle cut in the kinematic decomposition), where BH growth is dominated by accretion, which outpaces stellar mass growth from $z \sim 5$ to 0.

To explain why for the \texttt{BRAHMA} boxes, the \msigma evolution does not simply follow the top-down trend of the \mmstar evolution, we show the \mstarsigma relation in the right column of Figure~\ref{fig:MEGA}. 
The seed models have a negligible effect on the \mstarsigma relation, likely because BH accretion~(and therefore, AGN feedback) is too weak to have any significant influence on stellar mass assembly, especially in lower mass galaxies $\rm{M}_* \lesssim 10^9~M_{\odot}$. 
Importantly, regardless of the seed model, galaxies at fixed stellar mass exhibit higher velocity dispersions at higher redshifts. 
The key driver of this trend is that, at fixed $\rm{M}_*$, galaxies are more compact at higher redshifts (see Appendix~\ref{app:HMR} for more details). 
This evolution of \mstarsigma competes with the top-down \mmstar evolution, and as a result, prevents a top-down evolution in \msigma.

However, since there are no seed model variations in \mstarsigma, its evolution alone does not explain the seed model variations within the \msigma evolution. 
More specifically, it is not yet clear why we see negligible \msigma evolution for the most lenient seed model, and a bottom-up evolution~(at the high $\sigma$ end) for more restrictive seed models. 
To better understand the connection between the evolution of the \msigma relation and that of the \mmstar and \mstarsigma relations, we expand the time derivative of the median BH mass, $\bar{\rm{M}}_{\bullet}$ (in logarithmic units), at fixed $\sigma$ as

\begin{equation}
    \frac{\mathrm{d} \bar{\rm{M}}_{\bullet}}{\mathrm{d}t}\biggr|_\sigma = \frac{\partial \bar{\rm{M}}_{\bullet}}{\partial \bar{\rm{M}}_*}\biggr|_{t,\sigma} \cdot \frac{\mathrm{d} \bar{\rm{M}}_*}{\mathrm{d}t}\biggr|_\sigma + \frac{\partial \bar{\rm{M}}_{\bullet}}{\partial t}\biggr|_{\rm{M}_*,\sigma}
    \label{components}
\end{equation}

\begin{figure*}
    \centering
    \includegraphics[width=1.0\linewidth]{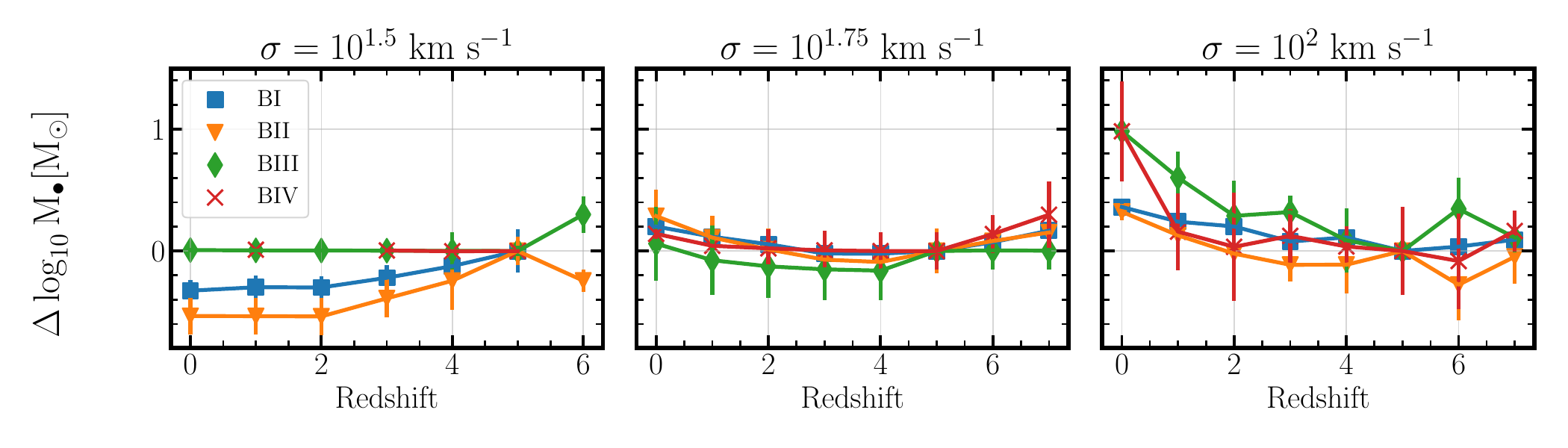}
    \caption{Redshift evolution of the median BH mass for the four \texttt{BRAHMA} simulations at fixed $\sigma$ values. All lines have been normalized to the median BH mass at $z=5$ to emphasize the different evolutions. Errorbars represent 95\% confidence intervals about the median mass calculated via boostrapping of the data. We clearly see here a negative redshift evolution for the more lenient seed models at $\sigma = 10^{1.5}$ \kms, very minimal redshift evolution for all seed models at $\sigma = 10^{1.5}$ \kms, and a stronger positive redshift evolution for more restrictive seed models at $\sigma = 10^2$ \kms. At $\sigma = 10^2$ \kms, for the more restrictive seed models, the increase in median BH mass occurs almost entirely at $z<2$.}
    \label{fig:msigma_redshift}

    \centering
    \includegraphics[width=1.0\linewidth]{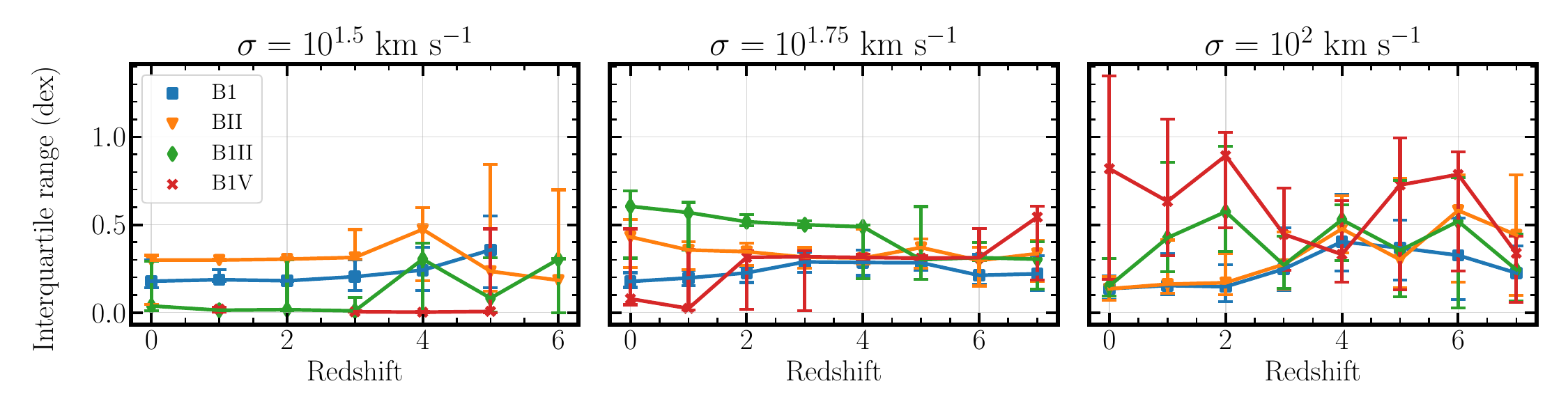}
    \caption{Redshift evolution in the scatter (shown using the interquartile range (IQR)) of the \msigma relation for subhalos of fixed $\sigma$. Errorbars again represent 95\% confidence intervals  calculated via boostrapping. With increasingly more restrictive seed models, the scatter and the variation in the scatter across all redshifts tends to increase for $\sigma = 10^2$ \kms, while the scatter remains very small at $\sigma = 10^{1.5}$ \kms for the more restrictive seed models.}
    \label{fig:msigma_scatter}
\end{figure*}

Here, $\bar{\rm{M}}_{\bullet}$ at fixed $\sigma$ is treated as a function of both stellar mass $\rm{M}_*$ and cosmic time $t$. 
In taking the total derivative $\mathrm{d} \bar{\rm{M}}_{\bullet}/\mathrm{d}t$, we are tracking the evolution of $\bar{\rm{M}}_{\bullet}$ across BH subsamples at fixed $\sigma$, but not at fixed $\rm{M}_*$. 
Figure~\ref{fig:msigma_redshift} shows this evolution of $\bar{\rm{M}}_{\bullet}$ across time for different $\sigma$ bins.
While this figure does not present fundamentally new information beyond what was shown in Figure~\ref{fig:MEGA}, it enables a closer examination of how the \msigma relation evolves as a function of $\sigma$ and seed model. 
In particular, we can clearly see that the evolution of $\bar{\rm{M}}_{\bullet}$ is increasingly steep for $\sigma = 100~\mathrm{km~s^{-1}}$ as the seed model becomes more restrictive. 
Notably, for the more restrictive seed models, almost all of this evolution occurs after $z \sim 2$.

To understand how these trends relate to the \mmstar and \mstarsigma evolution, we now consider the three components on the right-hand side (RHS) of Eq.~\ref{components}:

\begin{itemize}
    \item \textbf{Component 1} is $\frac{\mathrm{d} \bar{\rm{M}}_*}{\mathrm{d}t} \big|_\sigma$, which captures the evolution of stellar mass in galaxies within fixed-$\sigma$ subsamples. This essentially corresponds to the evolution of the \mstarsigma relation.
    
    \item \textbf{Component 2} is $\frac{\partial \bar{\rm{M}}_{\bullet}}{\partial \rm{M}_*} \big|_{t,\sigma}$, which measures the dependence of BH mass on stellar mass at fixed time and $\sigma$. As explained in Appendix~\ref{app:Confirming_msigma_redshift}~, this term can be reasonably approximated by the slope of the \mmstar relation.
    
    \item \textbf{Component 3} is $\frac{\partial \bar{\rm{M}}_{\bullet}}{\partial t} \big|_{\rm{M}_*, \sigma}$, which represents the explicit time evolution of $\bar{\rm{M}}_{\bullet}$ at fixed $\rm{M}_*$ and $\sigma$.
\end{itemize}

\begin{figure*}
    \centering

    \gridline{
        \fig{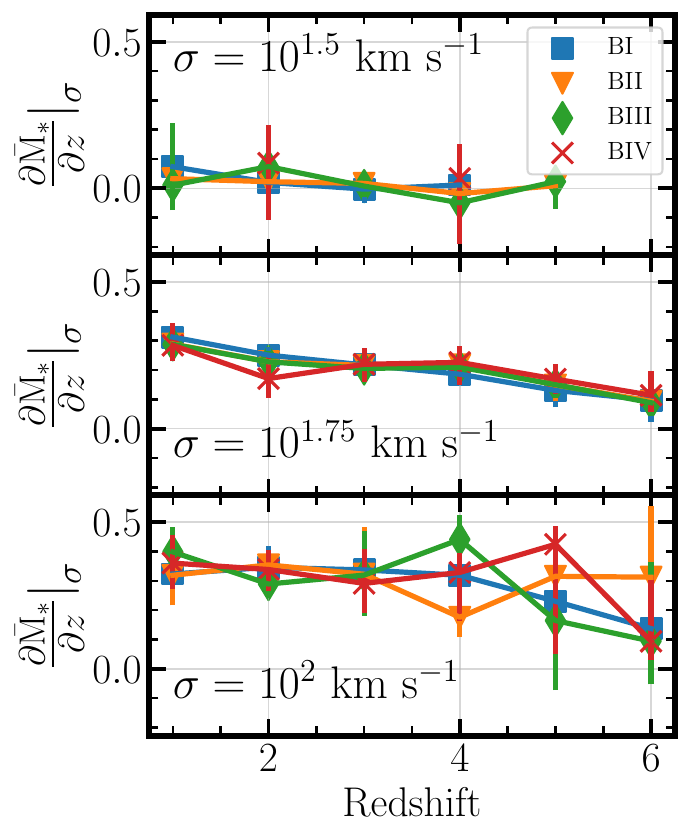}{0.45\textwidth}{(a) The change in stellar mass with decreasing redshift at fixed $\sigma$ values. The trends seen for each $\sigma$ value are very similar between \texttt{BRAHMA} boxes. Generally, $\mathrm{M}_*$ increases with decreasing redshift at fixed $\sigma$, with lower $\sigma$ values showing weaker redshift evolution.}
        \fig{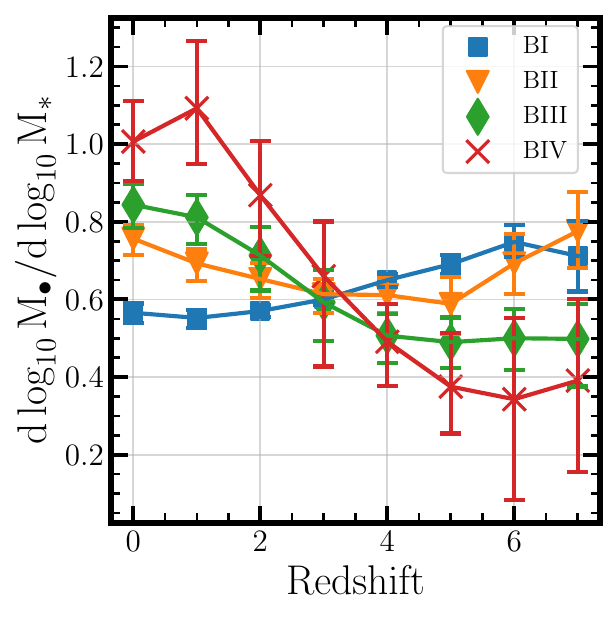}{0.45\textwidth}{(b) Redshift evolution of the slope of the \mmstar\ relation for each of the four \texttt{BRAHMA} simulations. At $z \sim 3$, there is a strong inversion in slope ordering: initially, more lenient seed models had larger slopes, whereas at later times, more restrictive models show much steeper slopes. This evolution is a key factor behind the changing \msigma\ relation.}
    }

    \vspace{0.5cm}

    \gridline{
        \fig{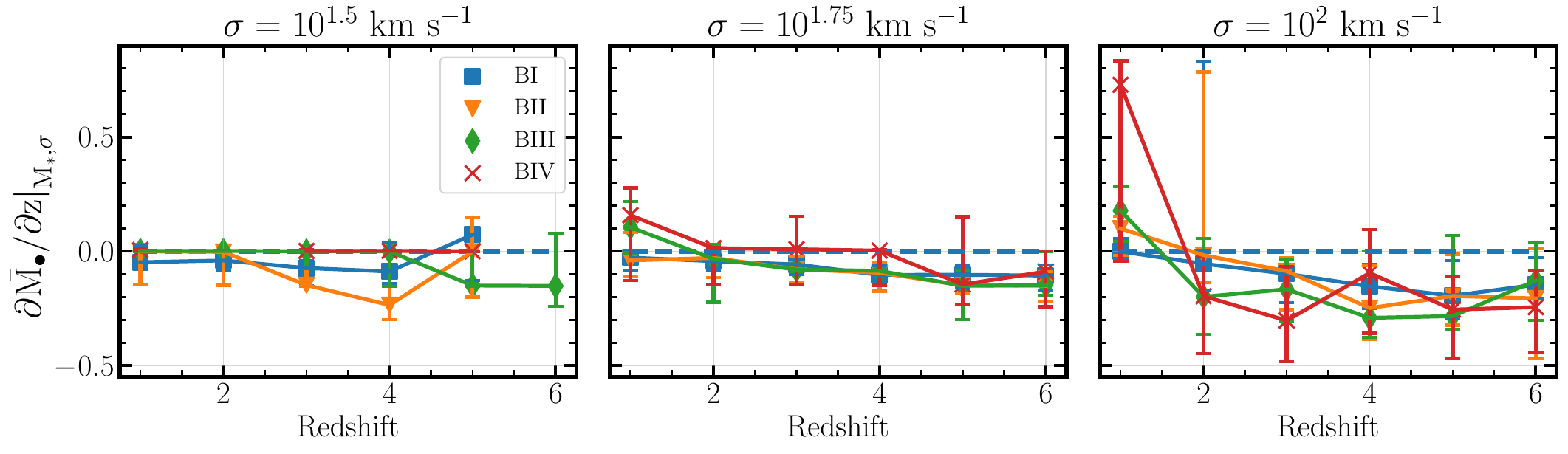}{0.9\textwidth}{(c) Explicit redshift dependence of median BH mass at constant $\sigma$ and $\mathrm{M}_*$. Highlighted in a blue dashed line is $\partial \mathrm{M_{\bullet}}/\partial z = 0$. For almost all redshifts for all seed models, the change in BH mass with redshift is negative. The sharp increase in $\partial \mathrm{M_{\bullet}}/\partial z$ at late times for the more restrictive models constitutes the other major contribution to a redshift-evolving \msigma\ relation.}
    }

    \caption{
        The evolution of the different components that determine the redshift evolution of the \msigma\ relation, as identified in Equation~\ref{components}. 
        Errorbars in each panel represent 95\% confidence intervals about the derivative being shown, obtained via bootstrapping of the data.}
    \label{fig:subfig_combined}
\end{figure*}

In Figure~\ref{fig:subfig_combined}, we study the interplay among the three components described above in determining the distinct evolution seen in the \msigma relation for different seed models. 
Note that we present the derivatives with respect to decreasing redshift (i.e., increasing cosmic time), as required by Eq.~\ref{components}. 

Component 1 (Figure \ref{fig:subfig_combined}a) shows negligible dependence on seeding, as also seen in column 3 of  Figure~\ref{fig:MEGA}. 
Therefore, differences between seed models in the \msigma evolution must originate from Components 2 and 3 (Figures~\ref{fig:subfig_combined}b and~\ref{fig:subfig_combined}c, respectively). 
Since component 1 is always positive, the relative signs of components 2 and 3 determine whether they cancel or reinforce each other, depending on the seed model. 
For both components, the dependence on redshift becomes stronger with more restrictive seeding. 
This also leads to significant differences in the interplay between components 2 and 3 across redshift for the different seed models. 
For the most lenient seed model, the combination of components 1 and 2 nearly cancels with component 3 at all redshifts, as components 1 and 2 remain positive and component 3 remains negative throughout (compare the blue line in Figure \ref{fig:subfig_combined}b with the leftmost panel in Figure~\ref{fig:subfig_combined}c), explaining why this seed model produces the weakest \msigma evolution. 
In contrast, for the most restrictive seed model, component 2 remains positive and increases toward lower redshift (red line in Figure\ref{fig:subfig_combined}b), while component 3 rises steeply around $z \sim 2$, transitioning from negative to positive for $\sigma = 10^{1.75}$ and $\sigma = 10^{2}~\mathrm{km~s^{-1}}$ (middle and right panels of Figure~\ref{fig:subfig_combined}c). 
As a result, the two components tend to cancel at $z \gtrsim 2$ but strongly reinforce each other at $z \lesssim 2$. 
This naturally explains the behavior for $\sigma=100$ \kms in Figure~\ref{fig:msigma_redshift} (rightmost panel), where the most restrictive seed model shows weak evolution in \msigma up to $z \sim 2$, followed by a steep increase between $z \sim 2$ and $z = 0$. 
All this also explains the evolution seen in the intermediate BII and BIII seed models, which are qualitatively similar to BIV but exhibit a less steep rise at $z \lesssim 2$. 
In Figure~\ref{fig:dmdz}, we explicitly confirm that these trends fully account for the evolution of $\rm{M}_{\bullet}$ at fixed $\sigma$.

\subsubsection{Physical drivers of the seed model variations in the \msigma evolution}

Now that we have isolated the contributions of Components~2 and~3 to the evolution of the median \msigma relation, we can investigate the implications of this analysis on the physics of BH growth. 
We begin with Component~2, which describes the evolution of the \mmstar relation. 
Figure~\ref{fig:subfig_combined}b shows how the slope of the \mmstar relation remains roughly constant with time for the lenient seed models and strongly steepens with time for the more restrictive seed models. 
As a result, the $z \lesssim 2$ slopes are higher for the restrictive seed models. 
All these trends are a natural consequence of the merger-dominated BH growth in low-mass ($\lesssim 10^9~M_{\odot}$) galaxies. 
In particular, as the seed models become more restrictive, BH-BH mergers are suppressed, and galaxy growth outpaces BH growth disproportionately more strongly in low-mass galaxies, compared to more massive galaxies where BHs can grow via accretion. 
We can readily see how the these processes steepen the \mmstar slope at $z \lesssim 2$ in the middle panels of Figure~\ref{fig:MEGA}.

We now discuss Component 3, which tracks how BH masses evolve with redshift at fixed $\rm{M}_*$ and $\sigma$. 
The fact that this component is predominantly negative at high redshift is a direct consequence of merger-driven BH growth, where galaxy growth naturally outpaces BH growth. 
In particular, because both $\rm{M}_*$ and $\sigma$ grow more rapidly than $\rm{M}_{\bullet}$, the latter appears to decrease with time when viewed at a fixed bin on the \mstarsigma plane. 
However, at $z \lesssim 2$, this component starts to increase and becomes positive by $z \sim 0$ for high $\sigma$~($\sigma \sim 100$ \kms). 
This is because in these high $\sigma \sim 100$ \kms and massive $\gtrsim 10^9~\rm{M}_{\odot}$ galaxies at $z\lesssim2$, accretion-driven BH growth accelerates and outpaces the growth in $\rm{M}_*$. 
Therefore, for the most restrictive BIV seed model, with the smallest amount of merger-driven growth, we see the strongest spike in component 3 at $z \lesssim 2$ for $\sigma = 100$~\kms in the rightmost panel of Fig. \ref{fig:subfig_combined}c.

Overall, we find that the evolution of the \msigma relation can be summarized as follows: The \msigma relation undergoes negligible evolution from $z\sim5-0$ for the most lenient BI seed model. 
For the most restrictive seed BIV model, there is a bottom-up evolution to $z\sim0$, with the majority of the evolution occurring for massive galaxies~($\gtrsim10^{9}~\rm{M}_{\odot}$) at high sigma end~($\sigma>100$ \kms) at $z\lesssim2$. 
These evolutionary trends are a consequence of the merger-dominated BH growth in low mass~($\lesssim10^{9}~\rm{M}_{\odot}$) galaxies and accretion dominated growth in  high mass~($\gtrsim10^{9}~\rm{M}_{\odot}$) galaxies, as identified in \cite{2025MNRAS.538..518B}.

\subsection{The redshift evolution in the scatter of the \msigma relation}

Having studied the median \msigma relation, we now study the redshift evolution of the scatter for different BH seed models. 
We quantify this scatter by the Interquartile Range~(IQR). 
Its evolution is plotted at fixed $\sigma$ in Fig.~\ref{fig:msigma_scatter}. 
Immediately, one can tell that the most lenient seed model has very little evolution in the scatter for all values of $\sigma$. 
However, as seed models become more restrictive, the overall scatter tends to increase for $\sigma=10^2$~\kms, while staying small for $\sigma = 10^{1.5}, 10^{1.75}$~\kms.

For low $\sigma=10^{1.5}$ \kms, the scatter becomes small for the two most restrictive seed models because the population of subhalos with this dispersion consist entirely of BHs that have grown very little past their initial seed mass. 
The galaxies that do see their BHs grow significantly have also all grown significantly in $\sigma$. 
Therefore, only the ungrown BHs are left in subhalos with $\sigma\sim10^{1.5}$ \kms.

At high $\sigma=100$~\kms, the larger scatter in the more restrictive seed models is also due to the stronger presence of ungrown BH seeds within more massive galaxies. 
More specifically, the increase in scatter is due to presence of \textit{both} grown as well as ungrown seeds in the more massive galaxies, which ``puffs up'' the \msigma relation as shown in Figure \ref{fig:Fullscatter}.
In contrast, for the more lenient seed models, almost all the seeds are able to readily grow and settle close to the local \msigma relation, leading to smaller scatter.

Unfortunately, our small volume significantly limits our ability to fully probe the \msigma relation. 
To that end, we show bootsrapped errors for the IQR as errorbars in Fig. \ref{fig:msigma_scatter}. 
Here, we clearly see these errorbars grow larger for more restrictive seed models, suggesting that there is some stochasticity in our current IQR values due to poor statistics. 
With a larger volume, we would expect this stochastic variation to be diminished, but the seed model variations in the scatter discussed in the previous paragraphs should remain.

\section{Discussion} \label{sec: Discussion}

\subsection{Implications from observational measurements of \msigma}\label{sec: Observations}

Having understood the evolution of the \msigma relation and its drivers, we now compare against currently available observations. 
At $z\sim0$, the seed model variations are most pronounced for our lower-mass BHs~($\rm{M}_{\bullet}\lesssim10^{7}~\rm{M}_{\odot}$) in galaxies with velocity dispersions $\sigma\sim50$--$80$ \kms. 
It is therefore useful to compare with observed local relations such as KH13, keeping in mind that these are based almost entirely on more massive BHs with $\rm{M}_{\bullet} \gtrsim 10^7 \mathrm{M}_\odot$. 
The intermediary BII and BIV seed models show reasonably good agreement with the ``extrapolated'' KH13 relation (Fig.~\ref{fig:msigmabinned}, right panel). 
The most lenient BI model yields a shallower slope and higher normalization than KH13 at $\sigma\sim50$--$80$ \kms, while the most restrictive BIV seed model produces lower normalizations. 
Similar conclusions arise when comparing the \mmstar relations with observations (see middle panels of Fig.~\ref{fig:MEGA}). 
For our massive BHs~($\gtrsim10^8~\rm{M}_{\odot}$) at $z=0$, the median \msigma relation shows a lower normalization compared to KH13. 
This is also the case for the \texttt{TNG} simulations. 
Since seed model variations are minimal for massive BHs at $z\sim0$, these discrepancies likely have stronger implications on the BH accretion and feedback models rather than the seed models, which is beyond the scope of this work.
 
At higher redshifts, we present the $z \sim 4-7$ JWST BLAGN on the \msigma plane, as measured by \cite{2024A&A...691A.145M} and \cite{2025arXiv250403551J}. 
Remarkably, these studies find that the high-z JWST AGN lie close to the local \msigma relation, yet appear significantly overmassive on the \mmstar plane.
This observation aligns strikingly well with our most lenient BI seed model, which produces similarly overmassive $z \sim 5$ BHs consistent with JWST data on the \mmstar relation, while simultaneously exhibiting negligible evolution in the \msigma relation. 
The BI seed model’s \msigma relation is also broadly consistent with the high-$z$ JWST AGN. 
Conversely, the more restrictive seed models, which predict bottom-up growth, systematically under-predict the \msigma relation compared to the JWST AGN.

Of course, uncertainties remain in the measurements of black hole mass, stellar mass, and stellar velocity dispersion for high-$z$ JWST AGN and their host galaxies. 
These include the possibility of BH mass overestimation~\citep{2025arXiv250316595R, 2025arXiv250316596N}, as well as potential underestimation of stellar masses due to uncertainties in modeling the AGN contribution~\citep{2020MNRAS.499.4325R}. 
Additionally, while stellar velocity dispersions are not directly measured at these redshifts, they are inferred from gas velocity dispersion measurements using a correction factor of 1.3 (see Section 5.1 of \citealt{2025arXiv250403551J}). 
Despite these uncertainties, our main conclusions regarding the impact of seeding on the \msigma relation and its evolution remain robust. 
Should these observational measurements stand the test of time, our most lenient seed model offers a natural explanation for the existence of high-$z$ BHs that appear overmassive relative to the local \mmstar relation, while still being entirely consistent with the local \msigma relation.

Another way to express the above results is to note that assembling the local \msigma relation as early as $z \gtrsim 5$—as suggested by JWST measurements—requires the abundant formation of heavy $\sim 10^5 \rm{M}_{\odot}$ seeds. 
Forming such seeds is challenging under canonical direct collapse black hole (DCBH) formation channels, which require extremely intense LW fluxes~($\gtrsim 1000J_{21}$; \cite{2010MNRAS.402.1249S,2014MNRAS.445..544S}). 
Under these flux conditions, the resulting seed abundances are more than two orders of magnitude lower than in our models~\citep{2025arXiv250200574O}.
Recent studies~\citep{2020OJAp....3E..15R, 2020MNRAS.492.3021R, 2019Natur.566...85W} suggest that dynamical heating in rapidly assembling halos may substantially lower the critical LW flux required for direct collapse. 
However, the resulting seed masses in these scenarios tend to be significantly lower, around $\sim 10^3 \rm{M}_{\odot}$. 
Alternative pathways for forming heavy seeds beyond the standard direct collapse path may exist—for example, rapid growth in Pop III star clusters~\citep{2025arXiv250507491V, 2025arXiv250320415R} or more exotic possibilities such as primordial black holes~\citep{2025JCAP...04..040Z} or self-interacting dark matter-aided collapse \citep{2025arXiv250400075S}. 
In any case, our results underscore the need for continued efforts to explore new mechanisms for efficiently forming heavy seeds using small-scale ``resolved physics" simulations.

\subsection{Modeling caveats}

As we found that the \msigma evolution is substantially influenced by the relative importance of BH growth via mergers versus accretion, it is important to acknowledge several caveats of the \texttt{BRAHMA} simulations that could affect this balance and must be considered when interpreting our results. 
One key caveat is our use of BH repositioning, which effectively causes BHs to merge instantaneously during galaxy mergers. 
This represents an optimistic scenario with minimal time delay between galaxy and BH mergers. 
\cite{2024MNRAS.533.1907B} studied the effect of merger delays on BH mass assembly and found that, for our lenient seeding models to reproduce the JWST observations, delays must be $\lesssim 750~\mathrm{Myr}$. 
We are currently investigating whether such timescales are possible under more accurate BH dynamics models that include unresolved dynamical friction~(Bhowmick et al., in prep).

If BH mergers turn out to be too inefficient to account for the high BH masses observed by JWST, we would need to enhance BH growth through accretion. 
In our simulations, Bondi-Hoyle accretion scales steeply with BH mass~($\rm{M}_{\bullet}^2$), which makes it difficult to grow low-mass BHs efficiently. 
Alternative accretion models such as the $\alpha$-disk model~\citep{1973A&A....24..337S}, the torque-limited model~\citep{2011MNRAS.415.1027H, 2017MNRAS.464.2840A}, and free-fall time–based models~\citep{2025arXiv250213241W} feature shallower mass dependence and could facilitate earlier growth of our seed BHs. 
However, in low-mass galaxies~($\lesssim 10^9 \rm{M}_{\odot}$), BH accretion is also suppressed by stellar feedback \citep{2017MNRAS.468.3935H, 2025arXiv250408035P}.
If stellar feedback evacuates gas from the BH’s vicinity, then accretion-driven growth would remain inefficient regardless of the chosen accretion prescription. 
In future work, we will systematically explore how the \msigma evolution and high-$z$ BH assembly respond to alternative BH accretion models and stellar feedback implementations.

\section{Conclusions} \label{sec: Conclusions}

In this work, we used the recently developed \texttt{BRAHMA} simulations \citep{2024MNRAS.533.1907B} to investigate whether different models for BH seed formation produce distinguishable features in BH-galaxy scaling relations, particularly the \msigma relation.
These $[18~\rm Mpc]^3$ \texttt{BRAHMA} boxes inherited the galaxy formation model of its predecessor \texttt{TNG}, except for the addition of new BH seeding prescriptions. 
These prescriptions initialized heavy ($\mathrm{M_{seed}} = 1.5 \times 10^5~\mathrm{M}_\odot$) seeds within halos based on a combination of dense and metal-poor gas content, local LW intensity, gas angular momentum (spin), and environmental richness. 
Each of these criteria was added cumulatively—in the order listed—across successive simulations, thereby making the seeding conditions progressively more restrictive. 
With our simulation volume and mass resolution~($\sim10^5~\rm{M}_{\odot}$ in gas), we were able to probe the \msigma relation in galaxies with $50\lesssim\sigma\lesssim200$ \kms at $z\sim5$, and  $50\lesssim\sigma\lesssim400$ \kms at $z=0$.

Our main results are as follows:

\begin{enumerate}
    \item At high redshift ($z > 2$), our BH seed models are distinguishable in the median \msigma relation across the full range of $\sigma$ values probed by our simulation volume. 
    This is because BH mass assembly is dominated by mergers within lower mass galaxies~($\lesssim10^{9}~M_{\odot}$).
    By $z \sim 0$, however, the seed models remain differentiable only in lower-mass galaxies with $50~\mathrm{km,s^{-1}} \lesssim \sigma \lesssim 80~\mathrm{km,s^{-1}}$, where merger-driven growth continues to dominate. 
    The differences between seed models become less pronounced in higher-mass~($\gtrsim10^9~M_{\odot}$) galaxies with $\sigma \gtrsim 80~\mathrm{km,s^{-1}}$, as BH growth is primarily fueled by gas accretion.
    
    \item Concurrently, the seed models exert a significant influence on the redshift evolution of the median \msigma\ relation. While \citet{2025MNRAS.538..518B} found that the evolution of \mmstar\ follows a predominantly top-down trend, the evolution of \msigma\ does not exhibit the same behavior. This is because the top-down evolution of \mmstar\ “competes” with the bottom-up evolution of \mstarsigma. The relative importance of these competing effects varies across seed models, driven by differences in the rates of merger-driven BH growth. Specifically,
    
    \begin{itemize}
        \item For the most lenient seed model, due to strong merger-dominated BH growth, there is no significant redshift evolution in the \msigma\ relation from $z \sim 5$ to $z \sim 0$. This is consistent with current JWST measurements of high-$z$ AGN, which suggest that these BHs appear overmassive relative to the local BHs on the \mmstar\ plane, but not on the \msigma\ plane.

        \item For the more restrictive seed models, the evolution in the \msigma\ relation remains modest at low $\sigma<50$ \kms, but exhibits a steep bottom-up trend at high $\sigma$ (e.g., $\sim 100$~\kms) between $z \sim 2$ and $z \sim 0$. This is driven by the onset of gas accretion as galaxies grow to stellar masses $\gtrsim 10^9~\rm{M}_{\odot}$, fueling rapid BH growth in more massive systems.
        
    \end{itemize}

\item Finally, the scatter (IQR) in the \msigma\ relation at high $\sigma$ (e.g., $100$~\kms) is larger for the more restrictive seed models.
This is because, whereas the lenient seed models enable consistent BH growth as galaxies evolve, the restrictive seed models result in a mixed population of grown and ungrown BH seeds, leading to greater diversity in BH masses at fixed $\sigma$.
  
\end{enumerate}

Our results highlight the potential for constraining BH seed models using the emerging population of high-$z$ JWST AGNs and their locations on the \msigma\ and \mmstar\ planes. 
Accommodating current observations—though subject to significant uncertainties—appears to require the abundant formation of heavy ($\sim10^5~\rm{M}_{\odot}$) seeds that can also undergo mergers on relatively short timescales ($\lesssim 750~\rm Myr$), as suggested by \citet{2024MNRAS.533.1907B}. 
Alternatively, one could investigate mechanisms for substantially enhancing accretion-driven BH growth in low-mass galaxies ($\lesssim 10^9~\rm{M}_{\odot}$) beyond what is seen in our simulations. 
These possibilities underscore the importance of continued efforts using small-scale “resolved physics” simulations to test the feasibility of these scenarios.

\section{Acknowledgements}

JK, AKB, PT, and AG acknowledge support from NSF-AST 2346977 and the NSF-Simons AI Institute for Cosmic Origins which is supported by the National Science Foundation under Cooperative Agreement 2421782 and the Simons Foundation award MPS-AI-00010515.
LB acknowledges support from NSF award AST-2307171.
The authors acknowledge Research Computing at The University of Virginia for providing computational resources and technical support that have contributed to the results reported within this publication. URL: https://rc.virginia.edu.

\appendix

\section{Explaining the redshift evolution of stellar mass at fixed \texorpdfstring{$\sigma$}{sigma}}\label{app:HMR}

\begin{figure}
    \centering
    \includegraphics[width=0.85\linewidth]{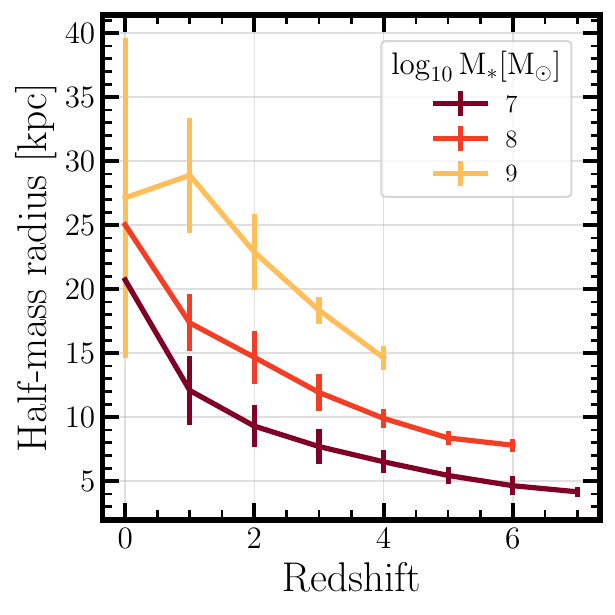}
    \caption{Redshift evolution of the half-mass radius of subhalos in the \texttt{BI} (most lenient model) for fixed stellar masses. The redshift evolution seen in the other three \texttt{BRAHMA} boxes is identical to that seen here, and so they are omitted to avoid redundancy. There is a strong increase in the size of subhalos with decreasing redshift, which explains the redshift evolution seen in Fig. \ref{fig:subfig_combined}a.}
    \label{fig:HMR_mstar}
\end{figure}

In order to explain for the redshift evolution of the \mstarsigma relation seen in Fig. \ref{fig:subfig_combined}a, we plot in Fig. \ref{fig:HMR_mstar} the redshift evolution of the median half-mass radius (HMR) of subhalos at fixed stellar masses. 
Note that we only plot the redshift evolution for the BI box, as the evolution here is identical between seed models, even more so than the redshift evolution seen in $\mathrm{M_*}$ for fixed $\sigma$. 
We see here that the HMR tends to increase with decreasing redshift for all stellar masses. 

This increase in size with decreasing redshift indicates that galaxies of a fixed mass were more compact at early times, a result that is in good agreement with current observations \citep{2008A&A...482...21C, 2017ApJ...839..127P, 2019ApJ...880...57M, 2025arXiv250104788C} which was also reproduced in \texttt{TNG} \citep{2018MNRAS.474.3976G}. 

This increase in HMR provides a excellent explanation for the redshift evolution seen in Fig. \ref{fig:subfig_combined}a. \cite{2010MNRAS.406.1220W} show that stellar velocity dispersion and galaxy size are related as follows:

\begin{equation}
    \sigma^2 \sim \frac{GM}{R_{\mathrm{HM}}}
\end{equation}

Based on this expression, we would expect the stellar mass to increase with increasing HMR for fixed $\sigma$, which is what exactly what we find.

\section{Confirming the explanation of the \texorpdfstring{\msigma}{msigma} redshift evolution}\label{app:Confirming_msigma_redshift}

\begin{figure*}
    \centering
    \includegraphics[width=1.0\linewidth]{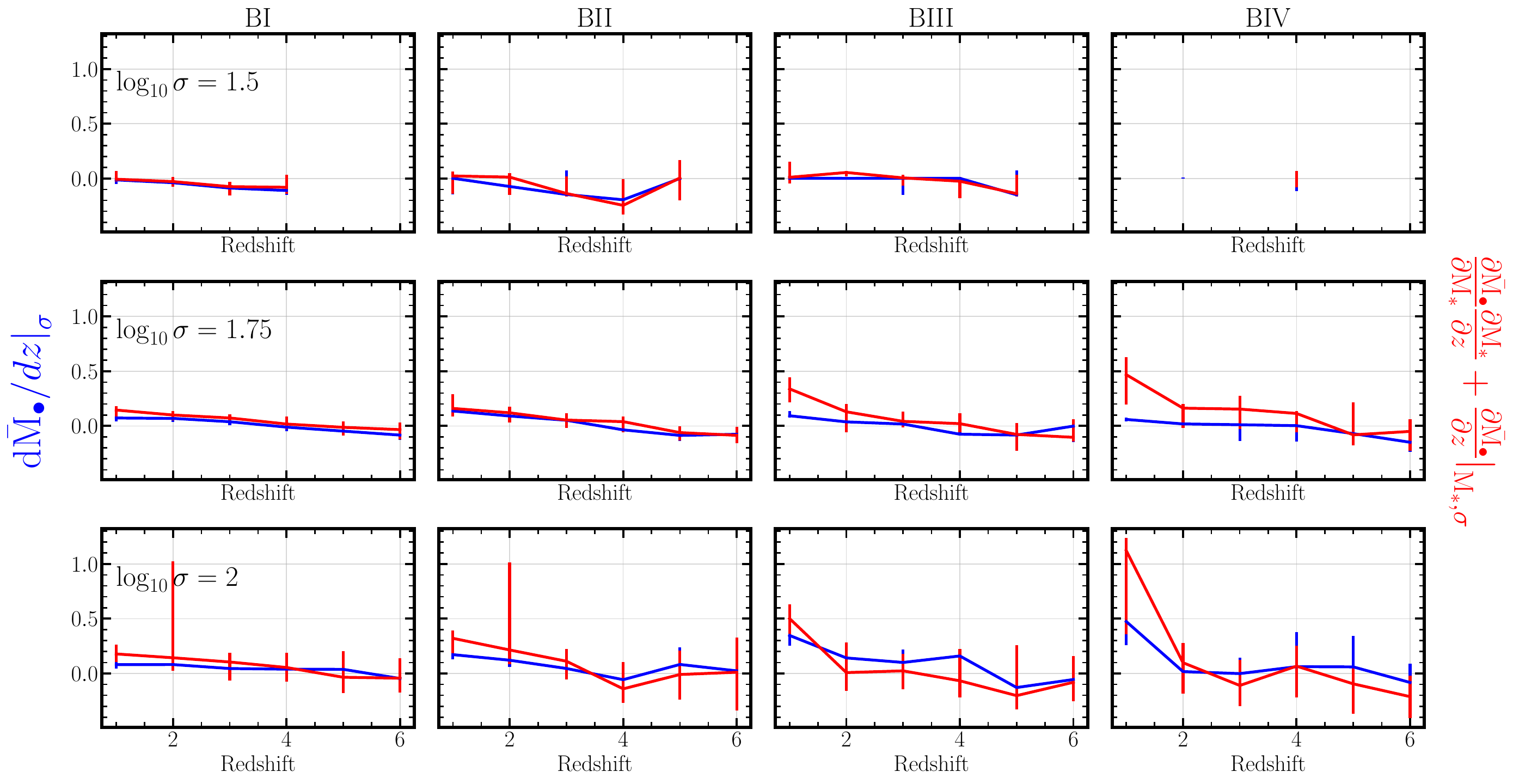}
    \caption{Approximated trends of the LHS (blue) and RHS (red) of Eqn. \ref{components} as a function of redshift for $\sigma = 10^{1.5},10^{1.75},10^2$ \kms. Even with the various approximations we made, the trends clearly follow the same redshift evolution, and show that a combination of the slope of the \mmstar relation, the stellar mass at constant $\sigma$ redshift evolution, and the explicit BH mass redshift dependence do a good job at explaining the redshift evolution in the \msigma relation.}
    \label{fig:dmdz}
\end{figure*}

In exploring the variations in the redshift evolution of the \msigma relation between different BH seed models, we made various approximations to calculate the different components of Eqn. \ref{components}.
In this appendix, we explain and justify the various approximations made and show that they produce reasonable results. 
To calculate derivatives of median values with respect to decreasing redshift, we use a finite difference approximation. 
For component 3, we only hold $\mathrm{M}_*$ constant for individual finite difference calculations, (i.e. for calculating $\Delta \mathrm{\bar{M}_\bullet}/\Delta z$ from $z=3$ to $z=1$ to approximate $\frac{\partial \mathrm{\bar{M}_\bullet}}{\partial z}(z=2)$), not across all redshifts. 
However, we do hold $\sigma$ constant across all redshifts, as necessitated by the left-hand side (LHS) of equation \ref{components}. 
For component 2, we approximate $\frac{\partial \mathrm{\bar{M_\bullet}}}{\partial\mathrm{M_*}}\big|_{z,\sigma}$ to be represented by the evolution of the \mmstar relation plotted in Figure \ref{fig:MEGA}~(middle panel), even though the latter did not assume a fixed $\sigma$.
While this effectively marginalizes $\frac{\partial \mathrm{\bar{M_\bullet}}}{\partial\mathrm{M_*}}\big|_{z,\sigma}$ over different $\sigma$ values at every $\rm{M}_*$, the approximation is nevertheless valid because of the relatively small scatter about the \mstarsigma relation~(the validity of this approximation is further corroborated by our comparison of the LHS and RHS of Eqn. \ref{components} in Fig. \ref{fig:dmdz}). 
Therefore, we perform a simple linear regression on the \mmstar data at each redshift for systems with BH masses greater than 5 times the seed mass\footnote{This mass cut was applied to mitigate the effect of the artificial flattening of the \mmstar relation for low-mass systems in our more restrictive models, which is a consequence of our fixed initial seed mass (see column 2 of Fig. \ref{fig:MEGA}). We thus calculate the slope for only those BHs which are growing.}. 

To assess the validity of our approximations of Eqn. \ref{components}, we show in Fig. \ref{fig:dmdz}, our calculations for the LHS and RHS of this equation as a function of redshift for $\sigma = 10^{1.5}, 10^{1.75}, 10^2$ \kms. 
While the LHS and RHS curves don't perfectly overlap, even our rough estimate shows that the curves follow the same trends as a function of redshift, and do tend to strongly overlap. 
The LHS and RHS points that don't strongly overlap are almost all within each others' 95\% bootstrapped confidence intervals.
The top right panel has no data because the most restrictive seed model has very few subhalos at $\sigma=10^{1.5}$ \kms that have been seeded with a BH. 
With a larger volume, this gap in the data would likely be filled, and better statistics could help alleviate some of the differences seen between our LHS and RHS curves, especially for our more restrictive seed models. 
Additionally, because our redshift sampling was very coarse (we only calculated finite difference approximations at $z=1,2,3,4,5,6$), we should expect that our finite difference approximations do not perfectly capture the true derivatives.

\bibliography{sample631}{}

\begin{thebibliography}{}
\expandafter\ifx\csname natexlab\endcsname\relax\def\natexlab#1{#1}\fi
\providecommand{\url}[1]{\href{#1}{#1}}
\providecommand{\dodoi}[1]{doi:~\href{http://doi.org/#1}{\nolinkurl{#1}}}
\providecommand{\doeprint}[1]{\href{http://ascl.net/#1}{\nolinkurl{http://ascl.net/#1}}}
\providecommand{\doarXiv}[1]{\href{https://arxiv.org/abs/#1}{\nolinkurl{https://arxiv.org/abs/#1}}}

\bibitem[{{Abadi} {et~al.}(2003){Abadi}, {Navarro}, {Steinmetz}, \& {Eke}}]{2003ApJ...591..499A}
{Abadi}, M.~G., {Navarro}, J.~F., {Steinmetz}, M., \& {Eke}, V.~R. 2003, \apj, 591, 499, \dodoi{10.1086/375512}

\bibitem[{{Agarwal} {et~al.}(2013){Agarwal}, {Davis}, {Khochfar}, {Natarajan}, \& {Dunlop}}]{2013MNRAS.432.3438A}
{Agarwal}, B., {Davis}, A.~J., {Khochfar}, S., {Natarajan}, P., \& {Dunlop}, J.~S. 2013, \mnras, 432, 3438, \dodoi{10.1093/mnras/stt696}

\bibitem[{{Akins} {et~al.}(2023){Akins}, {Casey}, {Allen}, {Bagley}, {Dickinson}, {Finkelstein}, {Franco}, {Harish}, {Arrabal Haro}, {Ilbert}, {Kartaltepe}, {Koekemoer}, {Liu}, {Long}, {McCracken}, {Paquereau}, {Papovich}, {Pirzkal}, {Rhodes}, {Robertson}, {Shuntov}, {Toft}, {Yang}, {Barro}, {Bisigello}, {Buat}, {Champagne}, {Cooper}, {Costantin}, {de La Vega}, {Drakos}, {Faisst}, {Fontana}, {Fujimoto}, {Gillman}, {G{\'o}mez-Guijarro}, {Gozaliasl}, {Hathi}, {Hayward}, {Hirschmann}, {Holwerda}, {Jin}, {Kocevski}, {Kokorev}, {Lambrides}, {Lucas}, {Magdis}, {Magnelli}, {McKinney}, {Mobasher}, {P{\'e}rez-Gonz{\'a}lez}, {Rich}, {Seill{\'e}}, {Talia}, {Urry}, {Valentino}, {Whitaker}, {Yung}, {Zavala}, {Cosmos-Web Team}, \& {Ceers Team}}]{2023ApJ...956...61A}
{Akins}, H.~B., {Casey}, C.~M., {Allen}, N., {et~al.} 2023, \apj, 956, 61, \dodoi{10.3847/1538-4357/acef21}

\bibitem[{{Akins} {et~al.}(2024){Akins}, {Casey}, {Lambrides}, {Allen}, {Andika}, {Brinch}, {Champagne}, {Cooper}, {Ding}, {Drakos}, {Faisst}, {Finkelstein}, {Franco}, {Fujimoto}, {Gentile}, {Gillman}, {Gozaliasl}, {Harish}, {Hayward}, {Hirschmann}, {Ilbert}, {Kartaltepe}, {Kocevski}, {Koekemoer}, {Kokorev}, {Liu}, {Long}, {McCracken}, {McKinney}, {Onoue}, {Paquereau}, {Renzini}, {Rhodes}, {Robertson}, {Shuntov}, {Silverman}, {Tanaka}, {Toft}, {Trakhtenbrot}, {Valentino}, \& {Zavala}}]{2024arXiv240610341A}
{Akins}, H.~B., {Casey}, C.~M., {Lambrides}, E., {et~al.} 2024, arXiv e-prints, arXiv:2406.10341, \dodoi{10.48550/arXiv.2406.10341}

\bibitem[{{Ananna} {et~al.}(2024){Ananna}, {Bogd{\'a}n}, {Kov{\'a}cs}, {Natarajan}, \& {Hickox}}]{2024ApJ...969L..18A}
{Ananna}, T.~T., {Bogd{\'a}n}, {\'A}., {Kov{\'a}cs}, O.~E., {Natarajan}, P., \& {Hickox}, R.~C. 2024, \apjl, 969, L18, \dodoi{10.3847/2041-8213/ad5669}

\bibitem[{{Angl{\'e}s-Alc{\'a}zar} {et~al.}(2017){Angl{\'e}s-Alc{\'a}zar}, {Dav{\'e}}, {Faucher-Gigu{\`e}re}, {{\"O}zel}, \& {Hopkins}}]{2017MNRAS.464.2840A}
{Angl{\'e}s-Alc{\'a}zar}, D., {Dav{\'e}}, R., {Faucher-Gigu{\`e}re}, C.-A., {{\"O}zel}, F., \& {Hopkins}, P.~F. 2017, \mnras, 464, 2840, \dodoi{10.1093/mnras/stw2565}

\bibitem[{{Arca Sedda} {et~al.}(2023){Arca Sedda}, {Kamlah}, {Spurzem}, {Rizzuto}, {Naab}, {Giersz}, \& {Berczik}}]{2023MNRAS.526..429A}
{Arca Sedda}, M., {Kamlah}, A. W.~H., {Spurzem}, R., {et~al.} 2023, \mnras, 526, 429, \dodoi{10.1093/mnras/stad2292}

\bibitem[{{Barnes} \& {Hut}(1986)}]{1986Natur.324..446B}
{Barnes}, J., \& {Hut}, P. 1986, \nat, 324, 446, \dodoi{10.1038/324446a0}

\bibitem[{{Bhowmick} {et~al.}(2022{\natexlab{a}}){Bhowmick}, {Blecha}, {Torrey}, {Kelley}, {Vogelsberger}, {Nelson}, {Weinberger}, \& {Hernquist}}]{2022MNRAS.510..177B}
{Bhowmick}, A.~K., {Blecha}, L., {Torrey}, P., {et~al.} 2022{\natexlab{a}}, \mnras, 510, 177, \dodoi{10.1093/mnras/stab3439}

\bibitem[{{Bhowmick} {et~al.}(2024{\natexlab{a}}){Bhowmick}, {Blecha}, {Torrey}, {Weinberger}, {Kelley}, {Vogelsberger}, {Hernquist}, \& {Somerville}}]{2024MNRAS.529.3768B}
---. 2024{\natexlab{a}}, \mnras, 529, 3768, \dodoi{10.1093/mnras/stae780}

\bibitem[{{Bhowmick} {et~al.}(2021){Bhowmick}, {Blecha}, {Torrey}, {Kelley}, {Vogelsberger}, {Kosciw}, {Nelson}, {Weinberger}, \& {Hernquist}}]{2021MNRAS.507.2012B}
---. 2021, \mnras, 507, 2012, \dodoi{10.1093/mnras/stab2204}

\bibitem[{{Bhowmick} {et~al.}(2022{\natexlab{b}}){Bhowmick}, {Blecha}, {Ni}, {Di Matteo}, {Torrey}, {Kelley}, {Vogelsberger}, {Weinberger}, \& {Hernquist}}]{2022MNRAS.516..138B}
{Bhowmick}, A.~K., {Blecha}, L., {Ni}, Y., {et~al.} 2022{\natexlab{b}}, \mnras, 516, 138, \dodoi{10.1093/mnras/stac2238}

\bibitem[{{Bhowmick} {et~al.}(2024{\natexlab{b}}){Bhowmick}, {Blecha}, {Torrey}, {Kelley}, {Weinberger}, {Vogelsberger}, {Hernquist}, {Somerville}, \& {Evans}}]{2024MNRAS.531.4311B}
{Bhowmick}, A.~K., {Blecha}, L., {Torrey}, P., {et~al.} 2024{\natexlab{b}}, \mnras, 531, 4311, \dodoi{10.1093/mnras/stae1386}

\bibitem[{{Bhowmick} {et~al.}(2024{\natexlab{c}}){Bhowmick}, {Blecha}, {Torrey}, {Somerville}, {Kelley}, {Vogelsberger}, {Weinberger}, {Hernquist}, \& {Sivasankaran}}]{2024MNRAS.533.1907B}
---. 2024{\natexlab{c}}, \mnras, 533, 1907, \dodoi{10.1093/mnras/stae1819}

\bibitem[{{Bhowmick} {et~al.}(2025){Bhowmick}, {Blecha}, {Torrey}, {Somerville}, {Kelley}, {Weinberger}, {Vogelsberger}, {Hernquist}, {Natarajan}, {Kho}, \& {Di Matteo}}]{2025MNRAS.538..518B}
---. 2025, \mnras, 538, 518, \dodoi{10.1093/mnras/staf269}

\bibitem[{{Bogd{\'a}n} {et~al.}(2024){Bogd{\'a}n}, {Goulding}, {Natarajan}, {Kov{\'a}cs}, {Tremblay}, {Chadayammuri}, {Volonteri}, {Kraft}, {Forman}, {Jones}, {Churazov}, \& {Zhuravleva}}]{2024NatAs...8..126B}
{Bogd{\'a}n}, {\'A}., {Goulding}, A.~D., {Natarajan}, P., {et~al.} 2024, Nature Astronomy, 8, 126, \dodoi{10.1038/s41550-023-02111-9}

\bibitem[{{Bondi}(1952)}]{1952MNRAS.112..195B}
{Bondi}, H. 1952, \mnras, 112, 195, \dodoi{10.1093/mnras/112.2.195}

\bibitem[{{Bondi} \& {Hoyle}(1944)}]{1944MNRAS.104..273B}
{Bondi}, H., \& {Hoyle}, F. 1944, \mnras, 104, 273, \dodoi{10.1093/mnras/104.5.273}

\bibitem[{{Bottrell} {et~al.}(2017){Bottrell}, {Torrey}, {Simard}, \& {Ellison}}]{2017MNRAS.467.2879B}
{Bottrell}, C., {Torrey}, P., {Simard}, L., \& {Ellison}, S.~L. 2017, \mnras, 467, 2879, \dodoi{10.1093/mnras/stx276}

\bibitem[{{Chabrier}(2003)}]{2003PASP..115..763C}
{Chabrier}, G. 2003, \pasp, 115, 763, \dodoi{10.1086/376392}

\bibitem[{{Cimatti} {et~al.}(2008){Cimatti}, {Cassata}, {Pozzetti}, {Kurk}, {Mignoli}, {Renzini}, {Daddi}, {Bolzonella}, {Brusa}, {Rodighiero}, {Dickinson}, {Franceschini}, {Zamorani}, {Berta}, {Rosati}, \& {Halliday}}]{2008A&A...482...21C}
{Cimatti}, A., {Cassata}, P., {Pozzetti}, L., {et~al.} 2008, \aap, 482, 21, \dodoi{10.1051/0004-6361:20078739}

\bibitem[{{Clausen} {et~al.}(2025){Clausen}, {Momcheva}, {Whitaker}, {Cutler}, {Bezanson}, {Dunlop}, {Grogin}, {Koekemoer}, {McLeod}, {McLure}, {Miller}, {Nelson}, {van der Wel}, {Wake}, \& {Wuyts}}]{2025arXiv250104788C}
{Clausen}, M., {Momcheva}, I., {Whitaker}, K.~E., {et~al.} 2025, arXiv e-prints, arXiv:2501.04788, \dodoi{10.48550/arXiv.2501.04788}

\bibitem[{{Davis} {et~al.}(1985){Davis}, {Efstathiou}, {Frenk}, \& {White}}]{1985ApJ...292..371D}
{Davis}, M., {Efstathiou}, G., {Frenk}, C.~S., \& {White}, S.~D.~M. 1985, \apj, 292, 371, \dodoi{10.1086/163168}

\bibitem[{{Di Matteo} {et~al.}(2005){Di Matteo}, {Springel}, \& {Hernquist}}]{2005Natur.433..604D}
{Di Matteo}, T., {Springel}, V., \& {Hernquist}, L. 2005, \nat, 433, 604, \dodoi{10.1038/nature03335}

\bibitem[{{Du} {et~al.}(2020){Du}, {Ho}, {Debattista}, {Pillepich}, {Nelson}, {Zhao}, \& {Hernquist}}]{2020ApJ...895..139D}
{Du}, M., {Ho}, L.~C., {Debattista}, V.~P., {et~al.} 2020, \apj, 895, 139, \dodoi{10.3847/1538-4357/ab8fa8}

\bibitem[{{Du} {et~al.}(2019){Du}, {Ho}, {Zhao}, {Shi}, {Debattista}, {Hernquist}, \& {Nelson}}]{2019ApJ...884..129D}
{Du}, M., {Ho}, L.~C., {Zhao}, D., {et~al.} 2019, \apj, 884, 129, \dodoi{10.3847/1538-4357/ab43cc}

\bibitem[{{Dunn} {et~al.}(2018){Dunn}, {Bellovary}, {Holley-Bockelmann}, {Christensen}, \& {Quinn}}]{2018ApJ...861...39D}
{Dunn}, G., {Bellovary}, J., {Holley-Bockelmann}, K., {Christensen}, C., \& {Quinn}, T. 2018, \apj, 861, 39, \dodoi{10.3847/1538-4357/aac7c2}

\bibitem[{{Evans} {et~al.}(2025){Evans}, {Blecha}, \& {Bhowmick}}]{2025MNRAS.536.2783E}
{Evans}, A.~E., {Blecha}, L., \& {Bhowmick}, A.~K. 2025, \mnras, 536, 2783, \dodoi{10.1093/mnras/stae2735}

\bibitem[{{Ferrarese} \& {Merritt}(2000)}]{2000ApJ...539L...9F}
{Ferrarese}, L., \& {Merritt}, D. 2000, \apjl, 539, L9, \dodoi{10.1086/312838}

\bibitem[{{Fragione} {et~al.}(2022){Fragione}, {Kocsis}, {Rasio}, \& {Silk}}]{2022ApJ...927..231F}
{Fragione}, G., {Kocsis}, B., {Rasio}, F.~A., \& {Silk}, J. 2022, \apj, 927, 231, \dodoi{10.3847/1538-4357/ac5026}

\bibitem[{{Furtak} {et~al.}(2024){Furtak}, {Labb{\'e}}, {Zitrin}, {Greene}, {Dayal}, {Chemerynska}, {Kokorev}, {Miller}, {Goulding}, {de Graaff}, {Bezanson}, {Brammer}, {Cutler}, {Leja}, {Pan}, {Price}, {Wang}, {Weaver}, {Whitaker}, {Atek}, {Bogd{\'a}n}, {Charlot}, {Curtis-Lake}, {van Dokkum}, {Endsley}, {Feldmann}, {Fudamoto}, {Fujimoto}, {Glazebrook}, {Juneau}, {Marchesini}, {Maseda}, {Nelson}, {Oesch}, {Plat}, {Setton}, {Stark}, \& {Williams}}]{2024Natur.628...57F}
{Furtak}, L.~J., {Labb{\'e}}, I., {Zitrin}, A., {et~al.} 2024, \nat, 628, 57, \dodoi{10.1038/s41586-024-07184-8}

\bibitem[{{Gebhardt} {et~al.}(2000){Gebhardt}, {Bender}, {Bower}, {Dressler}, {Faber}, {Filippenko}, {Green}, {Grillmair}, {Ho}, {Kormendy}, {Lauer}, {Magorrian}, {Pinkney}, {Richstone}, \& {Tremaine}}]{2000ApJ...539L..13G}
{Gebhardt}, K., {Bender}, R., {Bower}, G., {et~al.} 2000, \apjl, 539, L13, \dodoi{10.1086/312840}

\bibitem[{{Genel} {et~al.}(2018){Genel}, {Nelson}, {Pillepich}, {Springel}, {Pakmor}, {Weinberger}, {Hernquist}, {Naiman}, {Vogelsberger}, {Marinacci}, \& {Torrey}}]{2018MNRAS.474.3976G}
{Genel}, S., {Nelson}, D., {Pillepich}, A., {et~al.} 2018, \mnras, 474, 3976, \dodoi{10.1093/mnras/stx3078}

\bibitem[{{Greene} {et~al.}(2024){Greene}, {Labbe}, {Goulding}, {Furtak}, {Chemerynska}, {Kokorev}, {Dayal}, {Volonteri}, {Williams}, {Wang}, {Setton}, {Burgasser}, {Bezanson}, {Atek}, {Brammer}, {Cutler}, {Feldmann}, {Fujimoto}, {Glazebrook}, {de Graaff}, {Khullar}, {Leja}, {Marchesini}, {Maseda}, {Matthee}, {Miller}, {Naidu}, {Nanayakkara}, {Oesch}, {Pan}, {Papovich}, {Price}, {van Dokkum}, {Weaver}, {Whitaker}, \& {Zitrin}}]{2024ApJ...964...39G}
{Greene}, J.~E., {Labbe}, I., {Goulding}, A.~D., {et~al.} 2024, \apj, 964, 39, \dodoi{10.3847/1538-4357/ad1e5f}

\bibitem[{{Habouzit} {et~al.}(2017){Habouzit}, {Volonteri}, \& {Dubois}}]{2017MNRAS.468.3935H}
{Habouzit}, M., {Volonteri}, M., \& {Dubois}, Y. 2017, \mnras, 468, 3935, \dodoi{10.1093/mnras/stx666}

\bibitem[{{Habouzit} {et~al.}(2019){Habouzit}, {Genel}, {Somerville}, {Kocevski}, {Hirschmann}, {Dekel}, {Choi}, {Nelson}, {Pillepich}, {Torrey}, {Hernquist}, {Vogelsberger}, {Weinberger}, \& {Springel}}]{2019MNRAS.484.4413H}
{Habouzit}, M., {Genel}, S., {Somerville}, R.~S., {et~al.} 2019, \mnras, 484, 4413, \dodoi{10.1093/mnras/stz102}

\bibitem[{{Hahn} \& {Abel}(2011)}]{2011MNRAS.415.2101H}
{Hahn}, O., \& {Abel}, T. 2011, \mnras, 415, 2101, \dodoi{10.1111/j.1365-2966.2011.18820.x}

\bibitem[{{Harikane} {et~al.}(2023){Harikane}, {Zhang}, {Nakajima}, {Ouchi}, {Isobe}, {Ono}, {Hatano}, {Xu}, \& {Umeda}}]{2023ApJ...959...39H}
{Harikane}, Y., {Zhang}, Y., {Nakajima}, K., {et~al.} 2023, \apj, 959, 39, \dodoi{10.3847/1538-4357/ad029e}

\bibitem[{{Heger} {et~al.}(2023){Heger}, {M{\"u}ller}, \& {Mandel}}]{2023arXiv230409350H}
{Heger}, A., {M{\"u}ller}, B., \& {Mandel}, I. 2023, arXiv e-prints, arXiv:2304.09350, \dodoi{10.48550/arXiv.2304.09350}

\bibitem[{{Ho}(2008)}]{2008ARA&A..46..475H}
{Ho}, L.~C. 2008, \araa, 46, 475, \dodoi{10.1146/annurev.astro.45.051806.110546}

\bibitem[{{Hopkins} \& {Quataert}(2011)}]{2011MNRAS.415.1027H}
{Hopkins}, P.~F., \& {Quataert}, E. 2011, \mnras, 415, 1027, \dodoi{10.1111/j.1365-2966.2011.18542.x}

\bibitem[{{Huang} {et~al.}(2018){Huang}, {Di Matteo}, {Bhowmick}, {Feng}, \& {Ma}}]{2018MNRAS.478.5063H}
{Huang}, K.-W., {Di Matteo}, T., {Bhowmick}, A.~K., {Feng}, Y., \& {Ma}, C.-P. 2018, \mnras, 478, 5063, \dodoi{10.1093/mnras/sty1329}

\bibitem[{{Inayoshi} {et~al.}(2020){Inayoshi}, {Visbal}, \& {Haiman}}]{2020ARA&A..58...27I}
{Inayoshi}, K., {Visbal}, E., \& {Haiman}, Z. 2020, \araa, 58, 27, \dodoi{10.1146/annurev-astro-120419-014455}

\bibitem[{{Jeon} {et~al.}(2023){Jeon}, {Liu}, {Bromm}, \& {Finkelstein}}]{2023MNRAS.524..176J}
{Jeon}, J., {Liu}, B., {Bromm}, V., \& {Finkelstein}, S.~L. 2023, \mnras, 524, 176, \dodoi{10.1093/mnras/stad1877}

\bibitem[{{Juod{\v{z}}balis} {et~al.}(2025){Juod{\v{z}}balis}, {Maiolino}, {Baker}, {Lake}, {Scholtz}, {D'Eugenio}, {Trefoloni}, {Isobe}, {Tacchella}, {Bunker}, {Carniani}, {Charlot}, {Jones}, {Parlanti}, {Perna}, {Rinaldi}, {Robertson}, {{\"U}bler}, {Venturi}, \& {Willott}}]{2025arXiv250403551J}
{Juod{\v{z}}balis}, I., {Maiolino}, R., {Baker}, W.~M., {et~al.} 2025, arXiv e-prints, arXiv:2504.03551, \dodoi{10.48550/arXiv.2504.03551}

\bibitem[{{Katz} {et~al.}(1996){Katz}, {Weinberg}, \& {Hernquist}}]{1996ApJS..105...19K}
{Katz}, N., {Weinberg}, D.~H., \& {Hernquist}, L. 1996, \apjs, 105, 19, \dodoi{10.1086/192305}

\bibitem[{{Kido} {et~al.}(2025){Kido}, {Ioka}, {Hotokezaka}, {Inayoshi}, \& {Irwin}}]{2025arXiv250506965K}
{Kido}, D., {Ioka}, K., {Hotokezaka}, K., {Inayoshi}, K., \& {Irwin}, C.~M. 2025, arXiv e-prints, arXiv:2505.06965.
\newblock \doarXiv{2505.06965}

\bibitem[{{Klessen} \& {Glover}(2023)}]{2023ARA&A..61...65K}
{Klessen}, R.~S., \& {Glover}, S. C.~O. 2023, \araa, 61, 65, \dodoi{10.1146/annurev-astro-071221-053453}

\bibitem[{{Kokorev} {et~al.}(2024){Kokorev}, {Caputi}, {Greene}, {Dayal}, {Trebitsch}, {Cutler}, {Fujimoto}, {Labb{\'e}}, {Miller}, {Iani}, {Navarro-Carrera}, \& {Rinaldi}}]{2024ApJ...968...38K}
{Kokorev}, V., {Caputi}, K.~I., {Greene}, J.~E., {et~al.} 2024, \apj, 968, 38, \dodoi{10.3847/1538-4357/ad4265}

\bibitem[{{Kormendy} \& {Ho}(2013)}]{2013ARA&A..51..511K}
{Kormendy}, J., \& {Ho}, L.~C. 2013, \araa, 51, 511, \dodoi{10.1146/annurev-astro-082708-101811}

\bibitem[{{Kormendy} \& {Richstone}(1995)}]{1995ARA&A..33..581K}
{Kormendy}, J., \& {Richstone}, D. 1995, \araa, 33, 581, \dodoi{10.1146/annurev.aa.33.090195.003053}

\bibitem[{{Kov{\'a}cs} {et~al.}(2024){Kov{\'a}cs}, {Bogd{\'a}n}, {Natarajan}, {Werner}, {Azadi}, {Volonteri}, {Tremblay}, {Chadayammuri}, {Forman}, {Jones}, \& {Kraft}}]{2024ApJ...965L..21K}
{Kov{\'a}cs}, O.~E., {Bogd{\'a}n}, {\'A}., {Natarajan}, P., {et~al.} 2024, \apjl, 965, L21, \dodoi{10.3847/2041-8213/ad391f}

\bibitem[{{Laplace} {et~al.}(2025){Laplace}, {Schneider}, \& {Podsiadlowski}}]{2025A&A...695A..71L}
{Laplace}, E., {Schneider}, F.~R.~N., \& {Podsiadlowski}, P. 2025, \aap, 695, A71, \dodoi{10.1051/0004-6361/202451077}

\bibitem[{{Lin} {et~al.}(2024){Lin}, {Wang}, {Fan}, {Cai}, {Champagne}, {Sun}, {Volonteri}, {Yang}, {Hennawi}, {Ba{\~n}ados}, {Barth}, {Eilers}, {Farina}, {Liu}, {Jin}, {Jun}, {Lupi}, {Kakiichi}, {Mazzucchelli}, {Onoue}, {Pan}, {Pizzati}, {Rojas-Ruiz}, {Schindler}, {Trakhtenbrot}, {Shen}, {Trebitsch}, {Zhuang}, {Endsley}, {Meyer}, {Li}, {Li}, {Pudoka}, {Tee}, {Wu}, \& {Zhang}}]{2024ApJ...974..147L}
{Lin}, X., {Wang}, F., {Fan}, X., {et~al.} 2024, \apj, 974, 147, \dodoi{10.3847/1538-4357/ad6565}

\bibitem[{{Lodato} \& {Natarajan}(2006)}]{2006MNRAS.371.1813L}
{Lodato}, G., \& {Natarajan}, P. 2006, \mnras, 371, 1813, \dodoi{10.1111/j.1365-2966.2006.10801.x}

\bibitem[{{Luo} {et~al.}(2020){Luo}, {Shlosman}, {Nagamine}, \& {Fang}}]{2020MNRAS.492.4917L}
{Luo}, Y., {Shlosman}, I., {Nagamine}, K., \& {Fang}, T. 2020, \mnras, 492, 4917, \dodoi{10.1093/mnras/staa153}

\bibitem[{{Magorrian} {et~al.}(1998){Magorrian}, {Tremaine}, {Richstone}, {Bender}, {Bower}, {Dressler}, {Faber}, {Gebhardt}, {Green}, {Grillmair}, {Kormendy}, \& {Lauer}}]{1998AJ....115.2285M}
{Magorrian}, J., {Tremaine}, S., {Richstone}, D., {et~al.} 1998, \aj, 115, 2285, \dodoi{10.1086/300353}

\bibitem[{{Maiolino} {et~al.}(2024{\natexlab{a}}){Maiolino}, {Scholtz}, {Witstok}, {Carniani}, {D'Eugenio}, {de Graaff}, {{\"U}bler}, {Tacchella}, {Curtis-Lake}, {Arribas}, {Bunker}, {Charlot}, {Chevallard}, {Curti}, {Looser}, {Maseda}, {Rawle}, {Rodr{\'\i}guez del Pino}, {Willott}, {Egami}, {Eisenstein}, {Hainline}, {Robertson}, {Williams}, {Willmer}, {Baker}, {Boyett}, {DeCoursey}, {Fabian}, {Helton}, {Ji}, {Jones}, {Kumari}, {Laporte}, {Nelson}, {Perna}, {Sandles}, {Shivaei}, \& {Sun}}]{2024Natur.627...59M}
{Maiolino}, R., {Scholtz}, J., {Witstok}, J., {et~al.} 2024{\natexlab{a}}, \nat, 627, 59, \dodoi{10.1038/s41586-024-07052-5}

\bibitem[{{Maiolino} {et~al.}(2024{\natexlab{b}}){Maiolino}, {Scholtz}, {Curtis-Lake}, {Carniani}, {Baker}, {de Graaff}, {Tacchella}, {{\"U}bler}, {D'Eugenio}, {Witstok}, {Curti}, {Arribas}, {Bunker}, {Charlot}, {Chevallard}, {Eisenstein}, {Egami}, {Ji}, {Jones}, {Lyu}, {Rawle}, {Robertson}, {Rujopakarn}, {Perna}, {Sun}, {Venturi}, {Williams}, \& {Willott}}]{2024A&A...691A.145M}
{Maiolino}, R., {Scholtz}, J., {Curtis-Lake}, E., {et~al.} 2024{\natexlab{b}}, \aap, 691, A145, \dodoi{10.1051/0004-6361/202347640}

\bibitem[{{Marconi} \& {Hunt}(2003)}]{2003ApJ...589L..21M}
{Marconi}, A., \& {Hunt}, L.~K. 2003, \apjl, 589, L21, \dodoi{10.1086/375804}

\bibitem[{{Marinacci} {et~al.}(2018){Marinacci}, {Vogelsberger}, {Pakmor}, {Torrey}, {Springel}, {Hernquist}, {Nelson}, {Weinberger}, {Pillepich}, {Naiman}, \& {Genel}}]{2018MNRAS.480.5113M}
{Marinacci}, F., {Vogelsberger}, M., {Pakmor}, R., {et~al.} 2018, \mnras, 480, 5113, \dodoi{10.1093/mnras/sty2206}

\bibitem[{{Milosavljevi{\'c}} {et~al.}(2009){Milosavljevi{\'c}}, {Couch}, \& {Bromm}}]{2009ApJ...696L.146M}
{Milosavljevi{\'c}}, M., {Couch}, S.~M., \& {Bromm}, V. 2009, \apjl, 696, L146, \dodoi{10.1088/0004-637X/696/2/L146}

\bibitem[{{Mowla} {et~al.}(2019){Mowla}, {van Dokkum}, {Brammer}, {Momcheva}, {van der Wel}, {Whitaker}, {Nelson}, {Bezanson}, {Muzzin}, {Franx}, {MacKenty}, {Leja}, {Kriek}, \& {Marchesini}}]{2019ApJ...880...57M}
{Mowla}, L.~A., {van Dokkum}, P., {Brammer}, G.~B., {et~al.} 2019, \apj, 880, 57, \dodoi{10.3847/1538-4357/ab290a}

\bibitem[{{Naidu} {et~al.}(2025){Naidu}, {Matthee}, {Katz}, {de Graaff}, {Oesch}, {Smith}, {Greene}, {Brammer}, {Weibel}, {Hviding}, {Chisholm}, {Labb\textbackslash'e}, {Simcoe}, {Witten}, {Atek}, {Baggen}, {Belli}, {Bezanson}, {Boogaard}, {Bose}, {Covelo-Paz}, {Dayal}, {Fudamoto}, {Furtak}, {Giovinazzo}, {Goulding}, {Gronke}, {Heintz}, {Hirschmann}, {Illingworth}, {Inoue}, {Johnson}, {Leja}, {Leonova}, {McConachie}, {Maseda}, {Natarajan}, {Nelson}, {Setton}, {Shivaei}, {Sobral}, {Stefanon}, {Tacchella}, {Toft}, {Torralba}, {van Dokkum}, {van der Wel}, {Volonteri}, {Walter}, {Wang}, \& {Watson}}]{2025arXiv250316596N}
{Naidu}, R.~P., {Matthee}, J., {Katz}, H., {et~al.} 2025, arXiv e-prints, arXiv:2503.16596, \dodoi{10.48550/arXiv.2503.16596}

\bibitem[{{Naiman} {et~al.}(2018){Naiman}, {Pillepich}, {Springel}, {Ramirez-Ruiz}, {Torrey}, {Vogelsberger}, {Pakmor}, {Nelson}, {Marinacci}, {Hernquist}, {Weinberger}, \& {Genel}}]{2018MNRAS.477.1206N}
{Naiman}, J.~P., {Pillepich}, A., {Springel}, V., {et~al.} 2018, \mnras, 477, 1206, \dodoi{10.1093/mnras/sty618}

\bibitem[{{Natarajan} {et~al.}(2017){Natarajan}, {Pacucci}, {Ferrara}, {Agarwal}, {Ricarte}, {Zackrisson}, \& {Cappelluti}}]{2017ApJ...838..117N}
{Natarajan}, P., {Pacucci}, F., {Ferrara}, A., {et~al.} 2017, \apj, 838, 117, \dodoi{10.3847/1538-4357/aa6330}

\bibitem[{{Natarajan} {et~al.}(2024){Natarajan}, {Pacucci}, {Ricarte}, {Bogd{\'a}n}, {Goulding}, \& {Cappelluti}}]{2024ApJ...960L...1N}
{Natarajan}, P., {Pacucci}, F., {Ricarte}, A., {et~al.} 2024, \apjl, 960, L1, \dodoi{10.3847/2041-8213/ad0e76}

\bibitem[{{Nelson} {et~al.}(2018){Nelson}, {Pillepich}, {Springel}, {Weinberger}, {Hernquist}, {Pakmor}, {Genel}, {Torrey}, {Vogelsberger}, {Kauffmann}, {Marinacci}, \& {Naiman}}]{2018MNRAS.475..624N}
{Nelson}, D., {Pillepich}, A., {Springel}, V., {et~al.} 2018, \mnras, 475, 624, \dodoi{10.1093/mnras/stx3040}

\bibitem[{{O'Brennan} {et~al.}(2025){O'Brennan}, {Regan}, {Brennan}, {McCaffrey}, {Wise}, {Visbal}, \& {Norman}}]{2025arXiv250200574O}
{O'Brennan}, H., {Regan}, J.~A., {Brennan}, J., {et~al.} 2025, arXiv e-prints, arXiv:2502.00574, \dodoi{10.48550/arXiv.2502.00574}

\bibitem[{{Pacucci} \& {Loeb}(2024)}]{2024ApJ...964..154P}
{Pacucci}, F., \& {Loeb}, A. 2024, \apj, 964, 154, \dodoi{10.3847/1538-4357/ad3044}

\bibitem[{{Pacucci} {et~al.}(2023){Pacucci}, {Nguyen}, {Carniani}, {Maiolino}, \& {Fan}}]{2023ApJ...957L...3P}
{Pacucci}, F., {Nguyen}, B., {Carniani}, S., {Maiolino}, R., \& {Fan}, X. 2023, \apjl, 957, L3, \dodoi{10.3847/2041-8213/ad0158}

\bibitem[{{Pakmor} {et~al.}(2011){Pakmor}, {Bauer}, \& {Springel}}]{2011MNRAS.418.1392P}
{Pakmor}, R., {Bauer}, A., \& {Springel}, V. 2011, \mnras, 418, 1392, \dodoi{10.1111/j.1365-2966.2011.19591.x}

\bibitem[{{Pakmor} {et~al.}(2016){Pakmor}, {Pfrommer}, {Simpson}, {Kannan}, \& {Springel}}]{2016MNRAS.462.2603P}
{Pakmor}, R., {Pfrommer}, C., {Simpson}, C.~M., {Kannan}, R., \& {Springel}, V. 2016, \mnras, 462, 2603, \dodoi{10.1093/mnras/stw1761}

\bibitem[{{Patel} {et~al.}(2017){Patel}, {Hong}, {Quadri}, {Holden}, \& {Williams}}]{2017ApJ...839..127P}
{Patel}, S.~G., {Hong}, Y.~X., {Quadri}, R.~F., {Holden}, B.~P., \& {Williams}, R.~J. 2017, \apj, 839, 127, \dodoi{10.3847/1538-4357/aa6bf4}

\bibitem[{{P{\'e}rez-Gonz{\'a}lez} {et~al.}(2024){P{\'e}rez-Gonz{\'a}lez}, {Barro}, {Rieke}, {Lyu}, {Rieke}, {Alberts}, {Williams}, {Hainline}, {Sun}, {Pusk{\'a}s}, {Annunziatella}, {Baker}, {Bunker}, {Egami}, {Ji}, {Johnson}, {Robertson}, {Rodr{\'\i}guez Del Pino}, {Rujopakarn}, {Shivaei}, {Tacchella}, {Willmer}, \& {Willott}}]{2024ApJ...968....4P}
{P{\'e}rez-Gonz{\'a}lez}, P.~G., {Barro}, G., {Rieke}, G.~H., {et~al.} 2024, \apj, 968, 4, \dodoi{10.3847/1538-4357/ad38bb}

\bibitem[{{Petersson} {et~al.}(2025){Petersson}, {Hirschmann}, {Tress}, {Farcy}, {Glover}, {Klessen}, {Naab}, {Partmann}, \& {Whitworth}}]{2025arXiv250408035P}
{Petersson}, J., {Hirschmann}, M., {Tress}, R.~G., {et~al.} 2025, arXiv e-prints, arXiv:2504.08035, \dodoi{10.48550/arXiv.2504.08035}

\bibitem[{{Pezzulli} {et~al.}(2016){Pezzulli}, {Valiante}, \& {Schneider}}]{2016MNRAS.458.3047P}
{Pezzulli}, E., {Valiante}, R., \& {Schneider}, R. 2016, \mnras, 458, 3047, \dodoi{10.1093/mnras/stw505}

\bibitem[{{Pillepich} {et~al.}(2018{\natexlab{a}}){Pillepich}, {Springel}, {Nelson}, {Genel}, {Naiman}, {Pakmor}, {Hernquist}, {Torrey}, {Vogelsberger}, {Weinberger}, \& {Marinacci}}]{2018MNRAS.473.4077P}
{Pillepich}, A., {Springel}, V., {Nelson}, D., {et~al.} 2018{\natexlab{a}}, \mnras, 473, 4077, \dodoi{10.1093/mnras/stx2656}

\bibitem[{{Pillepich} {et~al.}(2018{\natexlab{b}}){Pillepich}, {Nelson}, {Hernquist}, {Springel}, {Pakmor}, {Torrey}, {Weinberger}, {Genel}, {Naiman}, {Marinacci}, \& {Vogelsberger}}]{2018MNRAS.475..648P}
{Pillepich}, A., {Nelson}, D., {Hernquist}, L., {et~al.} 2018{\natexlab{b}}, \mnras, 475, 648, \dodoi{10.1093/mnras/stx3112}

\bibitem[{{Planck Collaboration} {et~al.}(2016){Planck Collaboration}, {Ade}, {Aghanim}, {Arnaud}, {Ashdown}, {Aumont}, {Baccigalupi}, {Banday}, {Barreiro}, {Bartlett}, {Bartolo}, {Battaner}, {Battye}, {Benabed}, {Beno{\^\i}t}, {Benoit-L{\'e}vy}, {Bernard}, {Bersanelli}, {Bielewicz}, {Bock}, {Bonaldi}, {Bonavera}, {Bond}, {Borrill}, {Bouchet}, {Boulanger}, {Bucher}, {Burigana}, {Butler}, {Calabrese}, {Cardoso}, {Catalano}, {Challinor}, {Chamballu}, {Chary}, {Chiang}, {Chluba}, {Christensen}, {Church}, {Clements}, {Colombi}, {Colombo}, {Combet}, {Coulais}, {Crill}, {Curto}, {Cuttaia}, {Danese}, {Davies}, {Davis}, {de Bernardis}, {de Rosa}, {de Zotti}, {Delabrouille}, {D{\'e}sert}, {Di Valentino}, {Dickinson}, {Diego}, {Dolag}, {Dole}, {Donzelli}, {Dor{\'e}}, {Douspis}, {Ducout}, {Dunkley}, {Dupac}, {Efstathiou}, {Elsner}, {En{\ss}lin}, {Eriksen}, {Farhang}, {Fergusson}, {Finelli}, {Forni}, {Frailis}, {Fraisse}, {Franceschi}, {Frejsel}, {Galeotta}, {Galli}, {Ganga}, {Gauthier}, {Gerbino}, {Ghosh}, {Giard},
  {Giraud-H{\'e}raud}, {Giusarma}, {Gjerl{\o}w}, {Gonz{\'a}lez-Nuevo}, {G{\'o}rski}, {Gratton}, {Gregorio}, {Gruppuso}, {Gudmundsson}, {Hamann}, {Hansen}, {Hanson}, {Harrison}, {Helou}, {Henrot-Versill{\'e}}, {Hern{\'a}ndez-Monteagudo}, {Herranz}, {Hildebrandt}, {Hivon}, {Hobson}, {Holmes}, {Hornstrup}, {Hovest}, {Huang}, {Huffenberger}, {Hurier}, {Jaffe}, {Jaffe}, {Jones}, {Juvela}, {Keih{\"a}nen}, {Keskitalo}, {Kisner}, {Kneissl}, {Knoche}, {Knox}, {Kunz}, {Kurki-Suonio}, {Lagache}, {L{\"a}hteenm{\"a}ki}, {Lamarre}, {Lasenby}, {Lattanzi}, {Lawrence}, {Leahy}, {Leonardi}, {Lesgourgues}, {Levrier}, {Lewis}, {Liguori}, {Lilje}, {Linden-V{\o}rnle}, {L{\'o}pez-Caniego}, {Lubin}, {Mac{\'\i}as-P{\'e}rez}, {Maggio}, {Maino}, {Mandolesi}, {Mangilli}, {Marchini}, {Maris}, {Martin}, {Martinelli}, {Mart{\'\i}nez-Gonz{\'a}lez}, {Masi}, {Matarrese}, {McGehee}, {Meinhold}, {Melchiorri}, {Melin}, {Mendes}, {Mennella}, {Migliaccio}, {Millea}, {Mitra}, {Miville-Desch{\^e}nes}, {Moneti}, {Montier}, {Morgante}, {Mortlock},
  {Moss}, {Munshi}, {Murphy}, {Naselsky}, {Nati}, {Natoli}, {Netterfield}, {N{\o}rgaard-Nielsen}, {Noviello}, {Novikov}, {Novikov}, {Oxborrow}, {Paci}, {Pagano}, {Pajot}, {Paladini}, {Paoletti}, {Partridge}, {Pasian}, {Patanchon}, {Pearson}, {Perdereau}, {Perotto}, {Perrotta}, {Pettorino}, {Piacentini}, {Piat}, {Pierpaoli}, {Pietrobon}, {Plaszczynski}, {Pointecouteau}, {Polenta}, {Popa}, {Pratt}, \& {Pr{\'e}zeau}}]{2016A&A...594A..13P}
{Planck Collaboration}, {Ade}, P.~A.~R., {Aghanim}, N., {et~al.} 2016, \aap, 594, A13, \dodoi{10.1051/0004-6361/201525830}

\bibitem[{{Portegies Zwart} \& {McMillan}(2002)}]{2002ApJ...576..899P}
{Portegies Zwart}, S.~F., \& {McMillan}, S. L.~W. 2002, \apj, 576, 899, \dodoi{10.1086/341798}

\bibitem[{{Ramos Padilla} {et~al.}(2020){Ramos Padilla}, {Ashby}, {Smith}, {Mart{\'\i}nez-Galarza}, {Beverage}, {Dietrich}, {Higuera-G.}, \& {Weiner}}]{2020MNRAS.499.4325R}
{Ramos Padilla}, A.~F., {Ashby}, M.~L.~N., {Smith}, H.~A., {et~al.} 2020, \mnras, 499, 4325, \dodoi{10.1093/mnras/staa2813}

\bibitem[{{Regan} \& {Volonteri}(2024)}]{2024OJAp....7E..72R}
{Regan}, J., \& {Volonteri}, M. 2024, The Open Journal of Astrophysics, 7, 72, \dodoi{10.33232/001c.123239}

\bibitem[{{Regan} {et~al.}(2020{\natexlab{a}}){Regan}, {Wise}, {O'Shea}, \& {Norman}}]{2020MNRAS.492.3021R}
{Regan}, J.~A., {Wise}, J.~H., {O'Shea}, B.~W., \& {Norman}, M.~L. 2020{\natexlab{a}}, \mnras, 492, 3021, \dodoi{10.1093/mnras/staa035}

\bibitem[{{Regan} {et~al.}(2020{\natexlab{b}}){Regan}, {Wise}, {Woods}, {Downes}, {O'Shea}, \& {Norman}}]{2020OJAp....3E..15R}
{Regan}, J.~A., {Wise}, J.~H., {Woods}, T.~E., {et~al.} 2020{\natexlab{b}}, The Open Journal of Astrophysics, 3, 15, \dodoi{10.21105/astro.2008.08090}

\bibitem[{{Reines} \& {Volonteri}(2015)}]{2015ApJ...813...82R}
{Reines}, A.~E., \& {Volonteri}, M. 2015, \apj, 813, 82, \dodoi{10.1088/0004-637X/813/2/82}

\bibitem[{{Reinoso} {et~al.}(2023){Reinoso}, {Klessen}, {Schleicher}, {Glover}, \& {Solar}}]{2023MNRAS.521.3553R}
{Reinoso}, B., {Klessen}, R.~S., {Schleicher}, D., {Glover}, S. C.~O., \& {Solar}, P. 2023, \mnras, 521, 3553, \dodoi{10.1093/mnras/stad790}

\bibitem[{{Reinoso} {et~al.}(2025){Reinoso}, {Latif}, \& {Schleicher}}]{2025arXiv250320415R}
{Reinoso}, B., {Latif}, M.~A., \& {Schleicher}, D.~R.~G. 2025, arXiv e-prints, arXiv:2503.20415, \dodoi{10.48550/arXiv.2503.20415}

\bibitem[{{Rusakov} {et~al.}(2025){Rusakov}, {Watson}, {Nikopoulos}, {Brammer}, {Gottumukkala}, {Harvey}, {Heintz}, {Nielsen}, {Sim}, {Sneppen}, {Vijayan}, {Adams}, {Austin}, {Conselice}, {Goolsby}, \& {Toft}}]{2025arXiv250316595R}
{Rusakov}, V., {Watson}, D., {Nikopoulos}, G.~P., {et~al.} 2025, arXiv e-prints, arXiv:2503.16595, \dodoi{10.48550/arXiv.2503.16595}

\bibitem[{Scoggins \& Haiman(2024)}]{10.1093/mnras/stae1449}
Scoggins, M.~T., \& Haiman, Z. 2024, Monthly Notices of the Royal Astronomical Society, 531, 4584, \dodoi{10.1093/mnras/stae1449}

\bibitem[{{Scoggins} {et~al.}(2023){Scoggins}, {Haiman}, \& {Wise}}]{2023MNRAS.519.2155S}
{Scoggins}, M.~T., {Haiman}, Z., \& {Wise}, J.~H. 2023, \mnras, 519, 2155, \dodoi{10.1093/mnras/stac3715}

\bibitem[{{Shakura} \& {Sunyaev}(1973)}]{1973A&A....24..337S}
{Shakura}, N.~I., \& {Sunyaev}, R.~A. 1973, \aap, 24, 337

\bibitem[{{Shang} {et~al.}(2010){Shang}, {Bryan}, \& {Haiman}}]{2010MNRAS.402.1249S}
{Shang}, C., {Bryan}, G.~L., \& {Haiman}, Z. 2010, \mnras, 402, 1249, \dodoi{10.1111/j.1365-2966.2009.15960.x}

\bibitem[{{Shapiro} \& {Teukolsky}(1983)}]{1983bhwd.book.....S}
{Shapiro}, S.~L., \& {Teukolsky}, S.~A. 1983, {Black holes, white dwarfs and neutron stars. The physics of compact objects}, \dodoi{10.1002/9783527617661}

\bibitem[{{Shen} {et~al.}(2025){Shen}, {Shen}, {Xiao}, {Vogelsberger}, \& {Jiang}}]{2025arXiv250400075S}
{Shen}, T., {Shen}, X., {Xiao}, H., {Vogelsberger}, M., \& {Jiang}, F. 2025, arXiv e-prints, arXiv:2504.00075, \dodoi{10.48550/arXiv.2504.00075}

\bibitem[{{Smith} {et~al.}(2018){Smith}, {Regan}, {Downes}, {Norman}, {O'Shea}, \& {Wise}}]{2018MNRAS.480.3762S}
{Smith}, B.~D., {Regan}, J.~A., {Downes}, T.~P., {et~al.} 2018, \mnras, 480, 3762, \dodoi{10.1093/mnras/sty2103}

\bibitem[{{Soltan}(1982)}]{1982MNRAS.200..115S}
{Soltan}, A. 1982, \mnras, 200, 115, \dodoi{10.1093/mnras/200.1.115}

\bibitem[{{Spinoso} {et~al.}(2023){Spinoso}, {Bonoli}, {Valiante}, {Schneider}, \& {Izquierdo-Villalba}}]{2023MNRAS.518.4672S}
{Spinoso}, D., {Bonoli}, S., {Valiante}, R., {Schneider}, R., \& {Izquierdo-Villalba}, D. 2023, \mnras, 518, 4672, \dodoi{10.1093/mnras/stac3169}

\bibitem[{{Springel}(2010)}]{2010MNRAS.401..791S}
{Springel}, V. 2010, \mnras, 401, 791, \dodoi{10.1111/j.1365-2966.2009.15715.x}

\bibitem[{{Springel} \& {Hernquist}(2003)}]{2003MNRAS.339..289S}
{Springel}, V., \& {Hernquist}, L. 2003, \mnras, 339, 289, \dodoi{10.1046/j.1365-8711.2003.06206.x}

\bibitem[{{Springel} {et~al.}(2001){Springel}, {White}, {Tormen}, \& {Kauffmann}}]{2001MNRAS.328..726S}
{Springel}, V., {White}, S. D.~M., {Tormen}, G., \& {Kauffmann}, G. 2001, \mnras, 328, 726, \dodoi{10.1046/j.1365-8711.2001.04912.x}

\bibitem[{{Springel} {et~al.}(2018){Springel}, {Pakmor}, {Pillepich}, {Weinberger}, {Nelson}, {Hernquist}, {Vogelsberger}, {Genel}, {Torrey}, {Marinacci}, \& {Naiman}}]{2018MNRAS.475..676S}
{Springel}, V., {Pakmor}, R., {Pillepich}, A., {et~al.} 2018, \mnras, 475, 676, \dodoi{10.1093/mnras/stx3304}

\bibitem[{{Stark} {et~al.}(2025){Stark}, {Topping}, {Endsley}, \& {Tang}}]{2025arXiv250117078S}
{Stark}, D.~P., {Topping}, M.~W., {Endsley}, R., \& {Tang}, M. 2025, arXiv e-prints, arXiv:2501.17078, \dodoi{10.48550/arXiv.2501.17078}

\bibitem[{{Sugimura} {et~al.}(2014){Sugimura}, {Omukai}, \& {Inoue}}]{2014MNRAS.445..544S}
{Sugimura}, K., {Omukai}, K., \& {Inoue}, A.~K. 2014, \mnras, 445, 544, \dodoi{10.1093/mnras/stu1778}

\bibitem[{{Sun} {et~al.}(2025){Sun}, {Rieke}, {Lyu}, {Stone}, {Ji}, {Rinaldi}, {Willmer}, \& {Zhu}}]{2025arXiv250303675S}
{Sun}, Y., {Rieke}, G.~H., {Lyu}, J., {et~al.} 2025, arXiv e-prints, arXiv:2503.03675, \dodoi{10.48550/arXiv.2503.03675}

\bibitem[{{Tanaka} \& {Haiman}(2009)}]{2009ApJ...696.1798T}
{Tanaka}, T., \& {Haiman}, Z. 2009, \apj, 696, 1798, \dodoi{10.1088/0004-637X/696/2/1798}

\bibitem[{{Terrazas} {et~al.}(2020){Terrazas}, {Bell}, {Pillepich}, {Nelson}, {Somerville}, {Genel}, {Weinberger}, {Habouzit}, {Li}, {Hernquist}, \& {Vogelsberger}}]{2020MNRAS.493.1888T}
{Terrazas}, B.~A., {Bell}, E.~F., {Pillepich}, A., {et~al.} 2020, \mnras, 493, 1888, \dodoi{10.1093/mnras/staa374}

\bibitem[{{Toomre}(1964)}]{1964ApJ...139.1217T}
{Toomre}, A. 1964, \apj, 139, 1217, \dodoi{10.1086/147861}

\bibitem[{{Torrey} {et~al.}(2014){Torrey}, {Vogelsberger}, {Genel}, {Sijacki}, {Springel}, \& {Hernquist}}]{2014MNRAS.438.1985T}
{Torrey}, P., {Vogelsberger}, M., {Genel}, S., {et~al.} 2014, \mnras, 438, 1985, \dodoi{10.1093/mnras/stt2295}

\bibitem[{{Valiante} {et~al.}(2016){Valiante}, {Schneider}, {Volonteri}, \& {Omukai}}]{2016MNRAS.457.3356V}
{Valiante}, R., {Schneider}, R., {Volonteri}, M., \& {Omukai}, K. 2016, \mnras, 457, 3356, \dodoi{10.1093/mnras/stw225}

\bibitem[{{Vergara} {et~al.}(2025){Vergara}, {Askar}, {Kamlah}, {Spurzem}, {Flammini Dotti}, {Schleicher}, {Arca Sedda}, {Hypki}, {Giersz}, {Hurley}, {Berczik}, {Escala}, {Hoyer}, {Neumayer}, {Pang}, {Tanikawa}, {Cen}, \& {Naab}}]{2025arXiv250507491V}
{Vergara}, M.~C., {Askar}, A., {Kamlah}, A. W.~H., {et~al.} 2025, arXiv e-prints, arXiv:2505.07491, \dodoi{10.48550/arXiv.2505.07491}

\bibitem[{{Visbal} \& {Haiman}(2018)}]{2018ApJ...865L...9V}
{Visbal}, E., \& {Haiman}, Z. 2018, \apjl, 865, L9, \dodoi{10.3847/2041-8213/aadf3a}

\bibitem[{{Vogelsberger} {et~al.}(2013){Vogelsberger}, {Genel}, {Sijacki}, {Torrey}, {Springel}, \& {Hernquist}}]{2013MNRAS.436.3031V}
{Vogelsberger}, M., {Genel}, S., {Sijacki}, D., {et~al.} 2013, \mnras, 436, 3031, \dodoi{10.1093/mnras/stt1789}

\bibitem[{{Volonteri}(2010)}]{2010A&ARv..18..279V}
{Volonteri}, M. 2010, \aapr, 18, 279, \dodoi{10.1007/s00159-010-0029-x}

\bibitem[{{Volonteri} \& {Natarajan}(2009)}]{2009MNRAS.400.1911V}
{Volonteri}, M., \& {Natarajan}, P. 2009, \mnras, 400, 1911, \dodoi{10.1111/j.1365-2966.2009.15577.x}

\bibitem[{{Volonteri} \& {Rees}(2005)}]{2005ApJ...633..624V}
{Volonteri}, M., \& {Rees}, M.~J. 2005, \apj, 633, 624, \dodoi{10.1086/466521}

\bibitem[{{Weinberger} {et~al.}(2025){Weinberger}, {Bhowmick}, {Blecha}, {Bryan}, {Buchner}, {Hernquist}, {Hlavacek-Larrondo}, \& {Springel}}]{2025arXiv250213241W}
{Weinberger}, R., {Bhowmick}, A., {Blecha}, L., {et~al.} 2025, arXiv e-prints, arXiv:2502.13241, \dodoi{10.48550/arXiv.2502.13241}

\bibitem[{{Weinberger} {et~al.}(2020){Weinberger}, {Springel}, \& {Pakmor}}]{2020ApJS..248...32W}
{Weinberger}, R., {Springel}, V., \& {Pakmor}, R. 2020, \apjs, 248, 32, \dodoi{10.3847/1538-4365/ab908c}

\bibitem[{{Weinberger} {et~al.}(2017){Weinberger}, {Springel}, {Hernquist}, {Pillepich}, {Marinacci}, {Pakmor}, {Nelson}, {Genel}, {Vogelsberger}, {Naiman}, \& {Torrey}}]{2017MNRAS.465.3291W}
{Weinberger}, R., {Springel}, V., {Hernquist}, L., {et~al.} 2017, \mnras, 465, 3291, \dodoi{10.1093/mnras/stw2944}

\bibitem[{{Weinberger} {et~al.}(2018){Weinberger}, {Springel}, {Pakmor}, {Nelson}, {Genel}, {Pillepich}, {Vogelsberger}, {Marinacci}, {Naiman}, {Torrey}, \& {Hernquist}}]{2018MNRAS.479.4056W}
{Weinberger}, R., {Springel}, V., {Pakmor}, R., {et~al.} 2018, \mnras, 479, 4056, \dodoi{10.1093/mnras/sty1733}

\bibitem[{{Wiersma} {et~al.}(2009){Wiersma}, {Schaye}, \& {Smith}}]{2009MNRAS.393...99W}
{Wiersma}, R. P.~C., {Schaye}, J., \& {Smith}, B.~D. 2009, \mnras, 393, 99, \dodoi{10.1111/j.1365-2966.2008.14191.x}

\bibitem[{{Williams} {et~al.}(2024){Williams}, {Alberts}, {Ji}, {Hainline}, {Lyu}, {Rieke}, {Endsley}, {Suess}, {Sun}, {Johnson}, {Florian}, {Shivaei}, {Rujopakarn}, {Baker}, {Bhatawdekar}, {Boyett}, {Bunker}, {Cameron}, {Carniani}, {Charlot}, {Curtis-Lake}, {DeCoursey}, {de Graaff}, {Egami}, {Eisenstein}, {Gibson}, {Hausen}, {Helton}, {Maiolino}, {Maseda}, {Nelson}, {P{\'e}rez-Gonz{\'a}lez}, {Rieke}, {Robertson}, {Saxena}, {Tacchella}, {Willmer}, \& {Willott}}]{2024ApJ...968...34W}
{Williams}, C.~C., {Alberts}, S., {Ji}, Z., {et~al.} 2024, \apj, 968, 34, \dodoi{10.3847/1538-4357/ad3f17}

\bibitem[{{Wise} {et~al.}(2019){Wise}, {Regan}, {O'Shea}, {Norman}, {Downes}, \& {Xu}}]{2019Natur.566...85W}
{Wise}, J.~H., {Regan}, J.~A., {O'Shea}, B.~W., {et~al.} 2019, \nat, 566, 85, \dodoi{10.1038/s41586-019-0873-4}

\bibitem[{{Wolf} {et~al.}(2010){Wolf}, {Martinez}, {Bullock}, {Kaplinghat}, {Geha}, {Mu{\~n}oz}, {Simon}, \& {Avedo}}]{2010MNRAS.406.1220W}
{Wolf}, J., {Martinez}, G.~D., {Bullock}, J.~S., {et~al.} 2010, \mnras, 406, 1220, \dodoi{10.1111/j.1365-2966.2010.16753.x}

\bibitem[{{Ziparo} {et~al.}(2025){Ziparo}, {Gallerani}, \& {Ferrara}}]{2025JCAP...04..040Z}
{Ziparo}, F., {Gallerani}, S., \& {Ferrara}, A. 2025, \jcap, 2025, 040, \dodoi{10.1088/1475-7516/2025/04/040}

\end{thebibliography}
\bibliographystyle{aasjournal}

\end{document}